\documentclass[12pt]{article}
\usepackage[margin=0.95 in]{geometry}
\usepackage{amsmath} 
\usepackage{amssymb,amsfonts}
\usepackage[all]{xy}
\usepackage{graphicx}
\usepackage{simplewick}
\usepackage{amsmath}
\usepackage{amssymb}
\usepackage{float}
\usepackage{array}
\usepackage{tikz}
\usepackage{mathtools}
\usepackage{mathrsfs} 
\usepackage[hidelinks]{hyperref}
\usepackage{cite}
\numberwithin{equation}{section}
\usepackage[utf8x]{inputenc}

\newcommand{\be}{\begin{equation}}
\newcommand{\ee}{\end{equation}}
\def\bea{\begin{eqnarray}}
\def\eea{\end{eqnarray}}

\newcommand{\pa}{\partial}

\DeclareMathOperator{\Tr}{Tr}

\numberwithin{equation}{section}
\numberwithin{table}{section}

\begin{document}

\hypersetup{pageanchor=false}
\begin{titlepage}
\vbox{
    \halign{#\hfil         \cr
           } 
      }  
\vspace*{15mm}
\begin{center}
{\Large \bf Path Integrals on sl(2,R) Orbits}

\vspace*{10mm} 

{\large Sujay K. Ashok$^{a,b}$ and Jan Troost$^c$}
\vspace*{8mm}

$^a$The Institute of Mathematical Sciences, \\
		 IV Cross Road, C.I.T. Campus, \\
	 Taramani, Chennai, India 600113

\vspace{.6cm}

$^b$Homi Bhabha National Institute,\\ 
Training School Complex, Anushakti Nagar, \\
Mumbai, India 400094

\vspace{.6cm}

$^c$Laboratoire de Physique de l'\'Ecole Normale Sup\'erieure \\ 
 \hskip -.05cm
 CNRS, ENS, Universit\'e PSL,  Sorbonne Universit\'e, \\
 Universit\'e  Paris Cit\'e 
 \hskip -.05cm F-75005 Paris, France	 
\vspace*{0.8cm}
\end{center}

\begin{abstract}
We quantise orbits of  the adjoint group action on elements of the $sl(2,\mathbb{R})$  Lie algebra. The path integration along elliptic slices is akin to the coadjoint orbit quantization of compact Lie groups, and the calculation of the characters of elliptic group elements proceeds along the same lines as in compact groups. The computation of the trace of hyperbolic group elements in a diagonal basis as well as the calculation of the full group action on a hyperbolic basis  requires  considerably more technique. We determine the action of hyperbolic one-parameter subgroups of PSL$(2,\mathbb{R})$ on the adjoint orbits and discuss global subtleties in choices of adapted coordinate systems. Using the hyperbolic slicing of orbits, we describe the quantum mechanics of an irreducible $sl(2,\mathbb{R})$ representation in a hyperbolic basis and relate the basis to the mathematics of the  Mellin integral transform. We moreover discuss the representation theory of the double cover SL$(2,\mathbb{R})$ of PSL$(2,\mathbb{R})$ as well as that of its universal cover. Traces in the representations of these groups for both elliptic and hyperbolic elements are computed. Finally, we 
motivate our treatment of this elementary quantization problem by indicating  applications.

\end{abstract}

\end{titlepage}
\hypersetup{pageanchor=true}

\setcounter{tocdepth}{2}
\tableofcontents

\section{Introduction}
Many physical models have Lie groups  as their groups of symmetries. The energy eigenstates of these systems then transform in a representation of the Lie group. The representation is most often reducible.  It is a natural question to ask whether there are physical systems that give rise, after quantization, to a single irreducible representation of their symmetry group. An answer to this question is provided by the orbit method in mathematics \cite{Kostant,Souriau}. See e.g. \cite{Kirillov} for an introduction and overview. In physical terms, we can think of the orbit method as corresponding to the quantization of a system which has a phase space equal to the  orbit of a Lie algebra element under the (co-)adjoint action of the symmetry group. The phase space is then transitive and gives rise to an irreducible representation after quantization. In the physics literature, this idea has been applied in multiple contexts. An elementary application is to provide for a simple path integral realization of spin \cite{Souriau,Nielsen:1987sa,Johnson:1988qm}. This path integral quantization of orbits has been extended to large classes of representations of compact groups \cite{Alekseev:1988vx}. 
The mathematical orbit method is known to extend to classes of non-compact Lie groups \cite{Kirillov} and beyond. Non-compact Lie groups exhibit essentially new features. In this paper, we explore how one of these features influences the path integral approach to the quantization of adjoint orbits. 

We study the quantization of a point particle on the adjoint orbits inside the Lie algebra $sl(2,\mathbb{R})=so(2,1)=su(1,1)$. The Lie algebra contains six classes of orbits, depending on the reference Lie algebra element one conjugates. Moreover, exponentiating a multiple of a  non-trivial Lie algebra element can give rise to a one-parameter rotation subgroup, a boost subgroup or a scaling transformation (when the element has respectively negative, positive or zero Killing norm). 
In this paper we first review how the quantization of orbits that are sliced according to the action of an (elliptic) rotation subgroup remains rather close to the orbit theory of compact Lie groups. 
A new challenge and our main interest lies in understanding the path integral quantization when the orbit is sliced according to the action of a hyperbolic, boost one-parameter subgroup. This is a first example in which one clearly steps outside the realm of adjoint orbit theory for compact Lie groups.  The technicalities involved in performing the quantization are many, and we address some of them in detail. In mathematics, these technicalities lead to interesting results in special function theory and analysis \cite{PinkBook}. Physics problems also abound in which these one-parameter subgroups play an important role. The basic physics insights we provide in this paper are useful to address more ambitious problems. We will mention a few of those motivating goals in the concluding section. 

The paper kicks off in section \ref{EllipticQuantization} with a first description of the $sl(2,\mathbb{R})$  orbits and the manner to quantize them in close analogy with compact groups \cite{Vergne,Witten:1987ty}.
An elliptic $so(2)$ subalgebra plays the starring role in this section which is largely a review.  As in later sections, we compute traces of group elements in irreducible representations in the path integral approach. We  use these character calculations as a benchmark for our understanding of the path integral and the representations it engenders. In section \ref{HyperbolicGeometry} we describe the action of a hyperbolic one-parameter subgroup on the orbits, and introduce adapted coordinate systems. We carefully analyze the global geometry of the orbit since it is crucial in understanding the domain and the nature of the path integration.  The meat of the paper is the description of the path integral reconstruction of various kernels that capture how irreducible representations of groups with Lie algebra $sl(2,\mathbb{R})$ are represented on function spaces. In section \ref{DiscreteOrbits}, we obtain the kernels for the case of discrete orbits while in section \ref{ContinuousOrbits}, we present the path integral quantization of continuous orbits.  In section \ref{CoveringGroups}, we generalize our treatment of the representations of the group PSL$(2,\mathbb{R})=\mathrm{SO}(2,1)$  to those of its double cover SL$(2,\mathbb{R})$ as well as their universal cover $\widetilde{G}$.  In section \ref{Conclusions}, we provide conclusions and sketch how the classic mathematical physics problem we address feeds into other research activities. Since we study an elementary
and timeless model, we prefer to remain agnostic  for the larger part of the paper about where to apply the  insights we gain.

\section{\texorpdfstring{The $sl(2,\mathbb{R})$ coadjoint Orbit Theory}{The sl(2,R) Coadjoint Orbit Theory}}
\label{EllipticQuantization}
\label{Elliptic}
In this section we review $sl(2,\mathbb{R})$ coadjoint orbit theory. We describe the classes of orbits and their geometry. An orbit is a phase space with a standard symplectic form \cite{Kirillov}. It is transitive under the group action and the quantization of an orbit therefore gives rise to a single irreducible representation of the Lie algebra \cite{Kirillov}. For $sl(2,\mathbb{R})$, the orbit method was  described in  detail in \cite{Vergne}  (see also \cite{Witten:1987ty}). The path integral quantization of orbits of compact Lie groups was treated in \cite{Alekseev:1988vx} in some generality. The generalization of this path integral treatment to $sl(2,\mathbb{R})$ was described summarily in  \cite{Troost:2003ge,Troost:2012ck}. In compact groups, Cartan subgroups are tori. The compact path integral treatment is most easily extended to a parameterization of $sl(2,\mathbb{R})$ orbits that priviliges an elliptic $so(2)$ subalgebra. We will review this standard approach to the quantization of  the coadjoint orbits of $sl(2,\mathbb{R})$ in this section and add a few details. Thus, we slice orbits along the action of a one-parameter elliptic subgroup.  We  concentrate on computing the traces of elliptic group elements in the unitary irreducible representation to which the coadjoint orbit gives rise.

\subsection{A Word on Topology}
\label{Topology}
Before we study the orbits in the Lie algebra $sl(2,\mathbb{R})=so(2,1)$, it is useful to review properties of Lie groups with this Lie algebra. The Lie group SO$(2,1)$ is a first member of this family. It has a trivial center and it coincides with the group PSL$(2,\mathbb{R})$.  The Lie group   SL$(2,\mathbb{R})=\mathrm{SU}(1,1)$ has a center $\mathbb{Z}_2$ subgroup generated by minus the identity $(-e)$. When we divide SL$(2,\mathbb{R})$ by the center, we recuperate the group PSL$(2,\mathbb{R})=\mathrm{SL}(2,\mathbb{R})/\mathbb{Z}_2=\mathrm{SO}(2,1)$.  Clearly, the universal cover $\widetilde{G}$ of SL$(2,\mathbb{R})$  also covers PSL$(2,\mathbb{R})$. 
For both groups SL$(2,\mathbb{R})$ and PSL$(2,\mathbb{R})$, their first homotopy group $\Pi_1$ equals $\mathbb{Z}$.  
The universal covering group $\widetilde{G}$ has a center equal to $\mathbb{Z}$. 
There is an element $z$ in the center of the universal cover $\widetilde{G}$ which generates the central subgroup $\mathbb{Z} \subset \widetilde{G}$ such that $\widetilde{G}/\langle z^2 \rangle=\mathrm{SL}(2,\mathbb{R})$ and $\widetilde{G}/\langle z \rangle=\mathrm{PSL}(2,\mathbb{R})$. Inside the group SL$(2,\mathbb{R})$, and only there, we have that $z=-e$. All three groups share the same adjoint action on the Lie algebra $sl(2,\mathbb{R})$. The group SL$(2,\mathbb{R})$ has a standard $2 \times 2$ matrix representation. Indeed, it can be defined as the group of two-by-two matrices with real entries and unit determinant. The description of the covering group $\widetilde{G}$ is more involved. We will provide a detailed description in Section \ref{repsofcovers}.

\subsection{The Orbits}

\begin{figure}
\begin{center}
\includegraphics[scale=0.65]{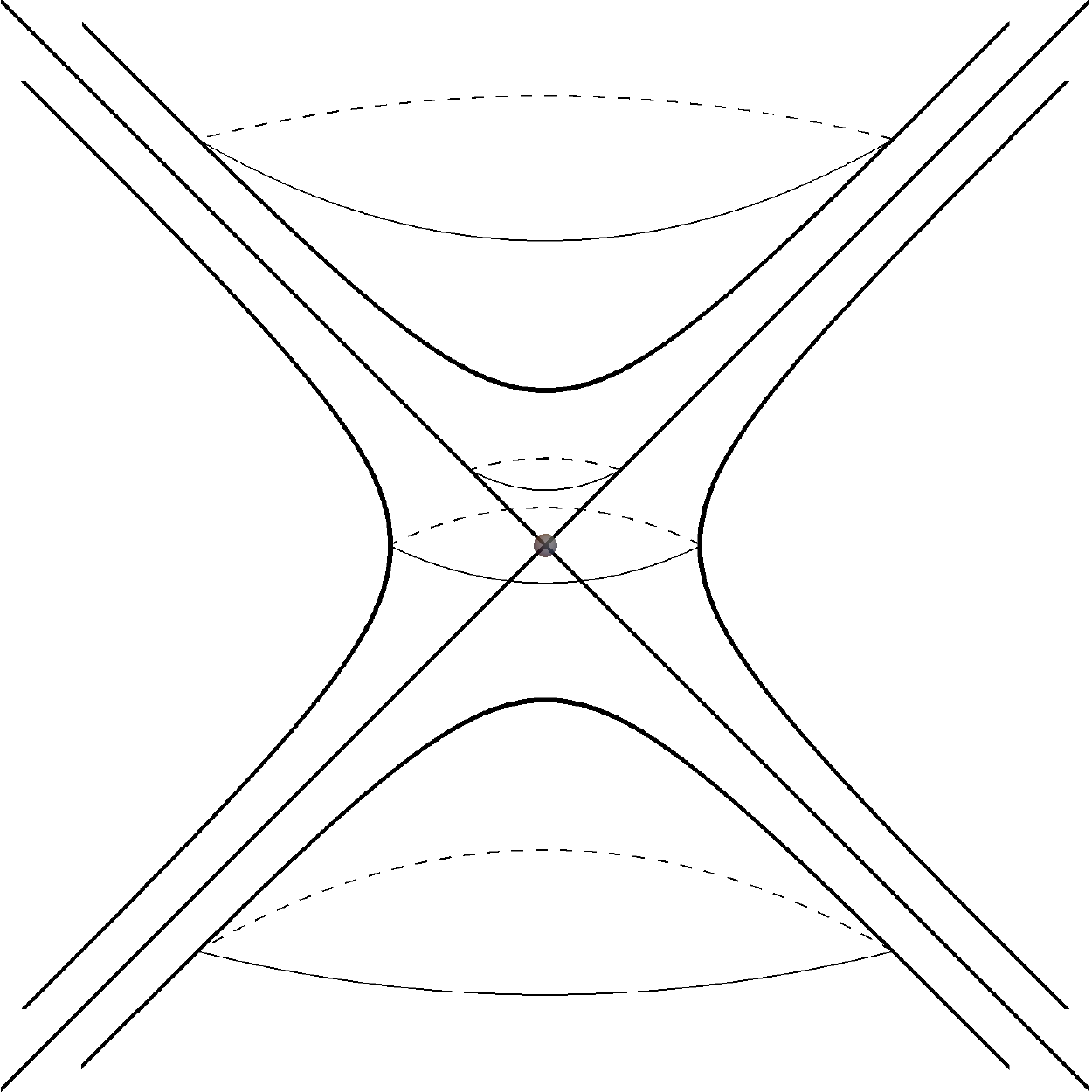}
\end{center}
\caption{The Orbits of $sl(2,\mathbb{R})$. The origin and the future and past light-cone are three unique orbits. There are future and past discrete orbits of time-like Lie algebra elements as well as continuous orbits of space-like Lie algebra elements. }
\label{SL2ROrbits}
\end{figure}
There are six types of orbits in the Lie algebra of  $sl(2,\mathbb{R})=so(2,1)$. See figure \ref{SL2ROrbits}. The Lie algebra has signature $(-,+,+)$ in the Killing metric and we can model the algebra as a three-dimensional  Minkowski vector space $\mathbb{R}^{1,2}$. There are three unique orbits which are the origin, the future light-cone and  the past light-cone. There are three more types of orbits with an infinite set of members: the continuous orbit that arises from conjugating a space-like element and the future and past time-like discrete orbits. Although the other cases can also be treated along the lines of our analysis with mild modifications, we concentrate on the three generic classes of orbits.\footnote{See e.g. \cite{Troost:2012ck} for a discussion of a light-cone orbit in the spirit of this section.}

\subsubsection{The Discrete Orbits}
Consider a coordinate system $(Y_0,Y_1,Y_2)$ on the Lie algebra such that the Killing norm of the corresponding Lie algebra element equals $-(Y_0)^2 + (Y_1)^2 + (Y_2)^2$.
The discrete orbits then satisfy the equation
\begin{equation}
-(Y_0)^2 + (Y_1)^2 + (Y_2)^2 = -\alpha^2 \, .
\end{equation}
There are two sheets that solve this equation. We consider each sheet separately. We will refer to the future sheet (with $Y_0>0$) as the discrete $D^+$ orbit  and the past sheet (with $Y_0<0$) as the discrete $D^-$ orbit. 
They can be parameterized by:
\begin{equation}
(Y_0,Y_1,Y_2) = \alpha (\pm \cosh r, \cos \phi \sinh r, \sin \phi \sinh r)
\, ,
\label{DiscreteOrbitRAndPhi}
\end{equation}
where the coordinates $(r,\phi)$ take values in the range $]-\infty,\infty[ \times [0,2 \pi[$. 
The orbit is a two-hyperboloid $H^2$ and equals the geometric coset SO$(2,1)/\mathrm{SO}(2)$. It is a phase space or symplectic manifold with  symplectic form $\Omega= \alpha \sinh r \, dr \wedge d \phi$.

\subsubsection{The Continuous Orbits}
The continuous orbits are the orbits of space-like elements $(Y_0,Y_1,Y_2)=(0,0,\alpha)$.  The orbit generated by group conjugation is the set of vectors that obey:
\begin{equation}
-(Y_0)^2 + (Y_1)^2 + (Y_2)^2 = \alpha^2 \, .
\end{equation}
We can parameterize the full space of solutions by 
\begin{equation}
(Y_0,Y_1,Y_2) = \alpha (\sinh r, \cos \phi \cosh r, \sin \phi \cosh r)
\, ,
\end{equation}
where the coordinates $(r,\phi)$ once more take  values in the domain $]-\infty,\infty[ \times [0,2 \pi[$. 
The space is a  $dS_2$ space-time or a hyperboloid of one sheet. Its coset description is SO$(2,1)/\mathrm{SO}(1,1)$ and it has the non-trivial topology of $\mathbb{R} \times S^1$. The symplectic form is $\Omega=\alpha \cosh r \, dr \wedge d \phi$. 

\subsection{The Elliptic Path Integral Quantization}
\label{ellipticquant}
\begin{figure}
\begin{center}
\includegraphics[scale=0.75]{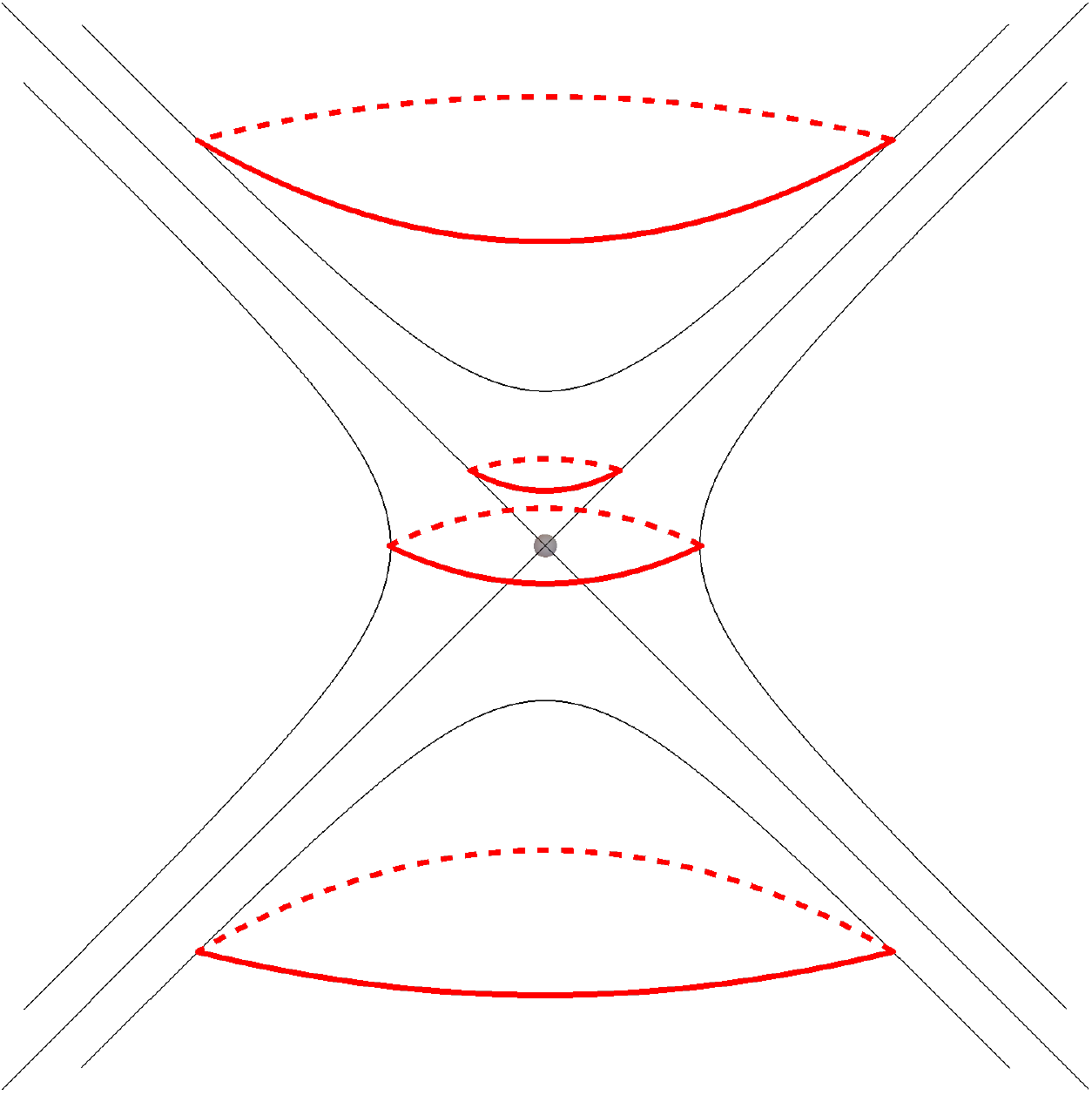}
\end{center}
\caption{An Elliptic Action on the Orbits of $sl(2,\mathbb{R})$.}
\label{EllipticActionOnSL2ROrbits}
\end{figure}

In this subsection, we review the path integral quantization of $sl(2,\mathbb{R})$ orbits. See e.g.  \cite{Vergne,Witten:1987ty} for original references. In the spirit of the path integral quantization of spin \cite{Nielsen:1987sa,Johnson:1988qm,Alekseev:1988vx}, we provide a physicist's viewpoint of the orbit method \cite{Kirillov}. We coalesce various  descriptions in the literature  into one unified discussion. As a guideline for our presentation, we concentrate on the calculation of  traces of elliptic group elements in the irreducible representation that arises upon quantization of the orbit. In other words, we evaluate the character of an irrep on an elliptic group element. To that end, we choose an adapted coordinate system on the orbits, and then compute the path integral for a particle travelling in a loop. 
The action of an elliptic element on the orbits is to rotate the orbit around the time axis. See Figure \ref{EllipticActionOnSL2ROrbits}. 
The parameterization of the orbits adapted to this action are precisely the coordinates we introduced previously. The coordinates are globally valid, up to one fixed point in discrete orbits. This renders the following analysis eminently tractable. 
The analysis  relies on the previous literature
\cite{Nielsen:1987sa,Johnson:1988qm,Alekseev:1988vx,Troost:2003ge,Troost:2012ck} -- in the interest of brevity of our review section, we omit  certain details that can be found in those references. 
\subsubsection{The Discrete Representations}
Our quantum mechanical system will be a particle whose phase space is the orbit. The action $S$ of the particle will be the primitive $d^{-1} \Omega$ of the symplectic form, pulled back to the world line $L$ of the particle, and integrated over the world line.
For the discrete plus orbit $D^+$, we found the symplectic form $\Omega=\alpha \sinh r dr \wedge d \phi$. Choosing $\alpha=j-1/2$, we can take the action to be:
\begin{equation}
S = (j-\frac{1}{2}) \int_L~dt~ \cosh r \, \dot{\phi} + \epsilon \int_L  \dot{\phi}
\, .
\label{DiscreteOrbitEllipticAction}
\end{equation}
We have added a total derivative term to the action with a coefficient $\epsilon$ that is arbitrary for now. It is easy to check that the phase space of our model equals the orbit. 
At this stage, the classical trajectories of the system  are static particles. Non-trivial dynamics is introduced if we introduce a Hamiltonian. We choose an elliptic Hamiltonian $H_{\text{ell}}$ proportional to the time-coordinate $Y_0$. The Hamiltonian equals the charge that generates rotations in the plane orthogonal to the time direction.
In the $(r,\phi)$ coordinates, the Hamiltonian only depends on the radius $r$:
\begin{align}
H_{\text{ell}}  &=    (j-\frac{1}{2}) \cosh r  \, .
\end{align}
The path integral of the quantum mechanical system we study equals:
\begin{equation}
K(\phi_f,\phi_i; j) = \int_{\phi_i}^{\phi_f} d \phi(t) \int d \eta(t) \, 
e^{iS + i \int dt H_{\text{ell}} } \, ,
\end{equation}
where the coordinate variable $\phi$ is chosen to have given initial and final conditions $(\phi_i,\phi_f)$ and the conjugate momentum variable 
\begin{equation}
\eta = \cosh r
\end{equation}
is freely integrated over in the appropriate range $[1,\infty[$. 
The path integral to be performed was discretized and discussed in \cite{Nielsen:1987sa,Johnson:1988qm,Alekseev:1988vx} in the compact $su(2)$ setting where the momentum integral $d \eta$ is over the interval $[-1,1]$. In the technical aspects and notation, we follow mostly \cite{Alekseev:1988vx}, except for a slight modification of how we regularize the path integral, discussed in Appendix B of \cite{Troost:2003ge}. We refer to those references for more detail. We compute a trace in the Hilbert space of the quantum mechanical system and therefore the path integral is over periodic configurations $\phi_i \equiv \phi_f$, namely loops. The coordinate $\phi$ is compact and it has periodicity $2 \pi$. Therefore, the loops can wind the compact direction and we compute the path integral over paths that satisfy:
\begin{equation}
\phi(T)=\phi(0) + 2 \pi w \, ,
\end{equation}
i.e. that wind the compact direction $w$ times in time $T$. 
The path integral is straightforwardly  computed \cite{Alekseev:1988vx}. One finds, after  performing the integration over almost all intermediate discretized positions and momenta \cite{Alekseev:1988vx}, the ordinary integral:
\begin{align}
\text{Tr}_j^+ \left( e^{ i T Y_0} \right) &= \sum_{w \in \mathbb{Z}} 
\int_{1}^{\infty} d \eta e^{2 \pi i w ( (j-\frac{1}{2}) \eta+\epsilon) + i   (j-\frac{1}{2}) \eta T}  \, .
\end{align}
Following  \cite{Troost:2003ge} Appendix B, we first perform the sum over the winding $w$ to obtain a Dirac comb:
\begin{align}
\text{Tr}_j^+ \left( e^{ iT Y_0} \right) &= \int_{1}^\infty d \eta \,  \delta_{\mathbb{Z}} ((j-\frac{1}{2}) \eta + \epsilon)
e^{ i   (j-\frac{1}{2}) \eta T}  \, .
\end{align}
The integration over the momentum $\eta$ picks out a particular subset of integers from the Dirac comb:
\begin{align}
\text{Tr}_j^+ \left( e^{ iT Y_0} \right) &= \sum_{m \in \mathbb{Z}, m \ge (j-\frac{1}{2})+\epsilon }
e^{ i   (m-\epsilon)  T}
\nonumber  \\
&= 
\frac{e^{ i \big(\lceil j-\frac{1}{2}+\epsilon\rceil-\epsilon \big)   T}}{1-e^{ i   T}} \, .
\end{align}
We used the ceiling function $\lceil{}\, \rceil$ to describe the result. We found the character of a discrete $D^{+}$ representation evaluated on an elliptic group element. From the path integral perspective, we can choose the parameter $\epsilon$ freely. The parameter $\epsilon$  acts as a shift of the conjugate momentum  and adds a phase to the overall result.  There is a choice of  the parameter $\epsilon$ such that the prefactor of the character precisely agrees with the standard character of the  $D^+_j$ representation of lowest weight $j$. We can choose $\epsilon$ such that it completes $j-1/2$ to a half-integer (by adding a  positive number smaller or equal than one). In that case we  find that $\lceil (j-1/2)+\epsilon \rceil=j+\epsilon$.  We  then have that the leading exponential has coefficient $e^{ i j   T}$. This is one choice of regularization scheme and it will be our preferred choice.\footnote{A feature of this scheme is that it agrees  with the regularization scheme for the finite representations in which the volume of phase space gives the dimension of the Hilbert space. See e.g. Appendix B of \cite{Troost:2003ge} for a discussion. The subtlety is a reflection of a zero point energy present in the quantum mechanical system. \label{VolumeFootnote}
} For discrete $D^-_j$ representations, the calculation of the character is analogous. 
Thus, our final result for the path integral equals the character:
\begin{align}
\text{Tr}_j^\pm \left( e^{ iT Y_0} \right)  &=  
\frac{e^{ \pm i j   T}}{1-e^{\pm i   T}} 
\, .
\end{align}
Indeed, the discrete orbits do give rise to individual irreducible discrete representations of $sl(2,\mathbb{R})$. 

We should note that our scheme
allows for the quantization of the orbit to provide representations of the group  PSL$(2,\mathbb{R})$, its universal cover, or an intermediate cover. One restricts the allowed values of the coefficient $j$ appropriately in order to find a spectrum for the elliptic generator that is consistent with the periodicity of the compact elliptic subgroup of the group in question. For PSL$(2,\mathbb{R})$ we have that the parameter $j$ needs to be integer, for SL$(2,\mathbb{R})$ it needs to be half-integer, and for the universal cover $\widetilde{G}$ it is positive. In all cases, we assumed that $j>1/2$. The quadratic Casimir of these representations is $c_2=-j(j-1)$. 

\begin{figure}
\begin{center}
\includegraphics[scale=0.5]{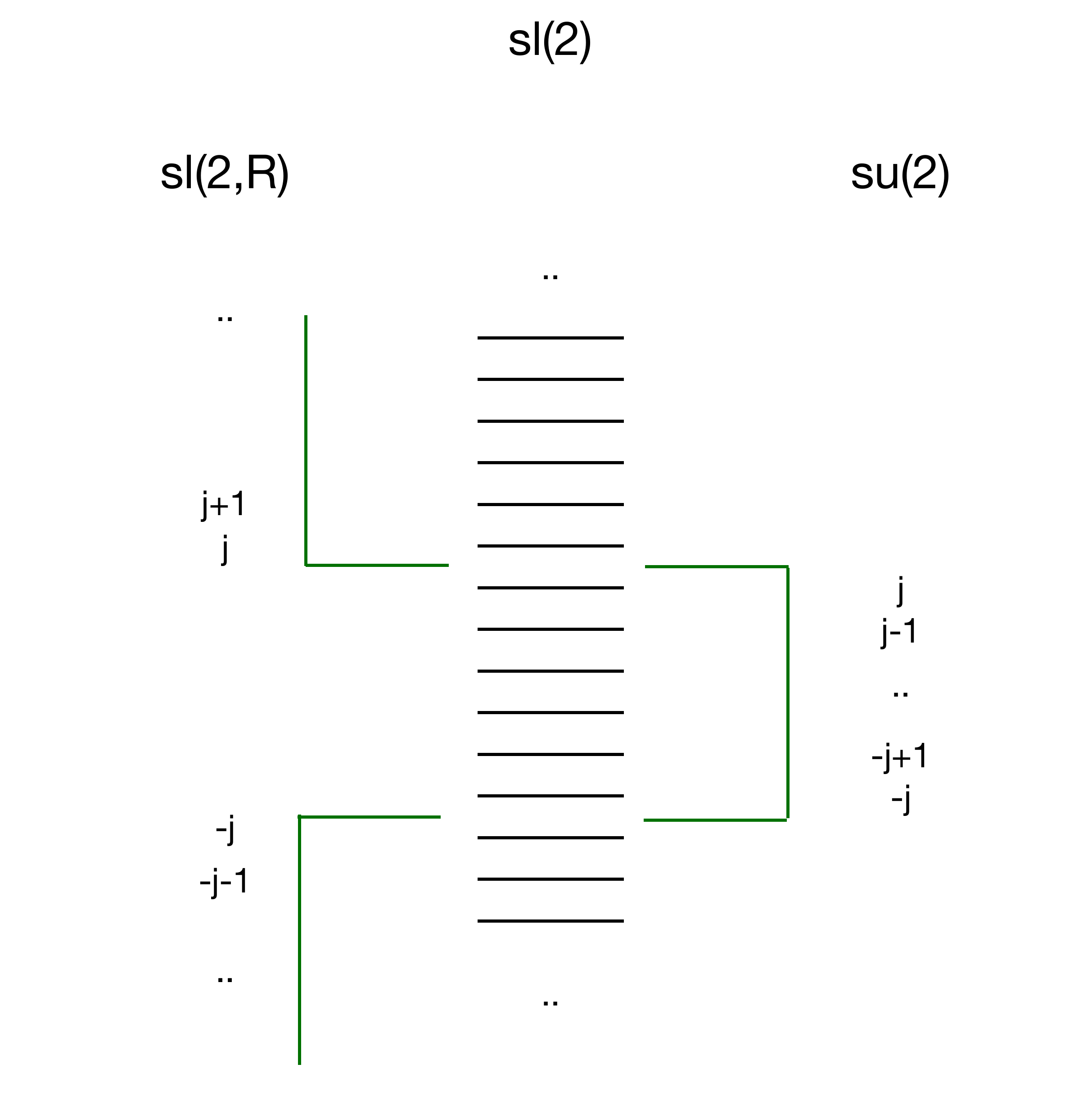}
\end{center}
\caption{Elliptic weight diagram for real and complex forms of the Lie algebra  $sl(2)=sl(2,\mathbb{C})$, including $sl(2,\mathbb{R})$ and $su(2)$. We assume $2j \in \mathbb{Z}$. In the left column standard conventions for $sl(2,\mathbb{R})$ are used and in the right column those for $su(2)$. Note that $j_{su(2)} = j_{sl(2,\mathbb{R})}-1$. The green brackets indicate slices of the tower of elliptic weights that are chosen for various $sl(2)$ representations. For $su(2)$, we have the finite dimensional representations of spin $j$ while for $sl(2,\mathbb{R})$ we have indicated discrete lowest and highest weight representations. The continuous representations of $sl(2,\mathbb{R})$ or $sl(2,\mathbb{C})$ contain a doubly infinite tower of states. }
\label{SL2EllipticWeightDiagram}
\end{figure}
We repeat that the calculation of the trace of an elliptic group element closely parallels the  calculation of the trace in an $su(2)$ representation \cite{Vergne,Alekseev:1988vx,Troost:2003ge}. Indeed, we can view the calculation as firstly projecting onto a tower of elliptic eigenvalues, and secondly, picking a slice of elliptic eigenvalues that is determined by the range of the momentum $\eta$. The difference between discrete representations of SL$(2,\mathbb{R})$ and finite representations of $su(2)$ lies then in the difference in the range of $\eta$ (determined by a trigonometric or a hyperbolic function). 
See Figure \ref{SL2EllipticWeightDiagram} for an illustration of these facts.\footnote{
Both types of representations can be unified in the larger framework of $sl(2,\mathbb{C})$ representation theory as discussed in \cite{Troost:2012ck}.  }

In summary, we have reviewed the calculation of  the trace of elliptic group elements in discrete representations of $sl(2,\mathbb{R})$ from the path integral quantization of the orbits.

\subsubsection{The Continuous Representations}
For continuous orbits, we find from the symplectic form $\Omega$ the action:
\begin{equation}
S = s \int~ dt~\sinh r \, \dot{\phi} -  \epsilon \int \dot{\phi}~.
\end{equation}
The integral over the momentum $\eta = \sinh r$ ranges from $-\infty$ to $+\infty$. The path integral calculation is very similar to the one performed before, and we wind up with the trace: 
\begin{align}
\text{Tr}_{\frac{1}{2}+is,\epsilon} \left(e^{i T Y_0} \right) &= \sum_{m \in \mathbb{Z}} e^{i   (m+\epsilon)T}
= e^{i   \epsilon T}  \delta_{\mathbb{Z}} (\frac{ T}{2 \pi}) 
\, .
\end{align}
For a non-trivial elliptic group element, the Dirac comb evaluates to zero. The final result is the character of a continuous representation with spin $j=1/2+is$ and intercept $\epsilon$. The restrictions on the parameter $\epsilon$ depend on the choice of covering group. For instance, for SL$(2,\mathbb{R})$, it is half-integer. The quadratic Casimir of these representations is $c_2=-j(j-1)=1/4+s^2$. 
Thus, we have calculated the trace of elliptic group elements in continuous representations through path integral means. So far, we have reviewed existing techniques in an intuitive manner and added a few minor extensions.

\section{\texorpdfstring{The Orbits of $sl(2,\mathbb{R})$ in Hyperbolic Slices}{The Orbits of SL(2,R) in Hyperbolic Slices}}
\label{HyperbolicGeometry}
There are three types of one-parameter subgroups of PSL$(2,\mathbb{R})$. Depending on whether the norm of the corresponding Lie algebra generator is negative, zero or positive, we call these subgroups elliptic, parabolic or hyperbolic. In the previous section, we worked in a parameterization of the orbits adapted to the action of an elliptic one-parameter subgroup. In the quantum mechanics, this is closely linked to a description of the Hilbert space in terms of a basis of states that diagonalise an elliptic generator of the Lie algebra.\footnote{Intuitively, one can think of each elliptic generator eigenstate as corresponding to an elliptic slice of area equal to Planck's constant $h$.} In compact groups, there are only elliptic generators. In non-compact groups however, there are more possibilities. It is well-known that this elementary fact gives rise to a  plethora of non-trivial technical results. One example  in which these extra possibilities are exploited is in the development of special function theory in relation to representation theory. A useful reference in this field, closely related to our technical developments below is \cite{PinkBook}. 

In this section, we  develop another perspective on the geometry of the discrete and continuous orbits. We slice the orbits according to the action of a hyperbolic one-parameter subgroup on the orbit. Moreover, we choose coordinates adapted to this slicing. We then study the path integral in these coordinates, and in particular, we compute the trace of hyperbolic elements of the group. In the chosen parameterization, the system will once again be free. There are other complications through. In the case of continuous orbits, the coordinate system will no longer be global, and this renders the path integral treatment more challenging. Before we path integrate in sections \ref{DiscreteOrbits} and  \ref{HyperbolicPathIntegral}, in this section we gain more insight into the hyperbolic geometry of our orbits. 

\begin{figure}
\begin{center}
\includegraphics[scale=0.8]{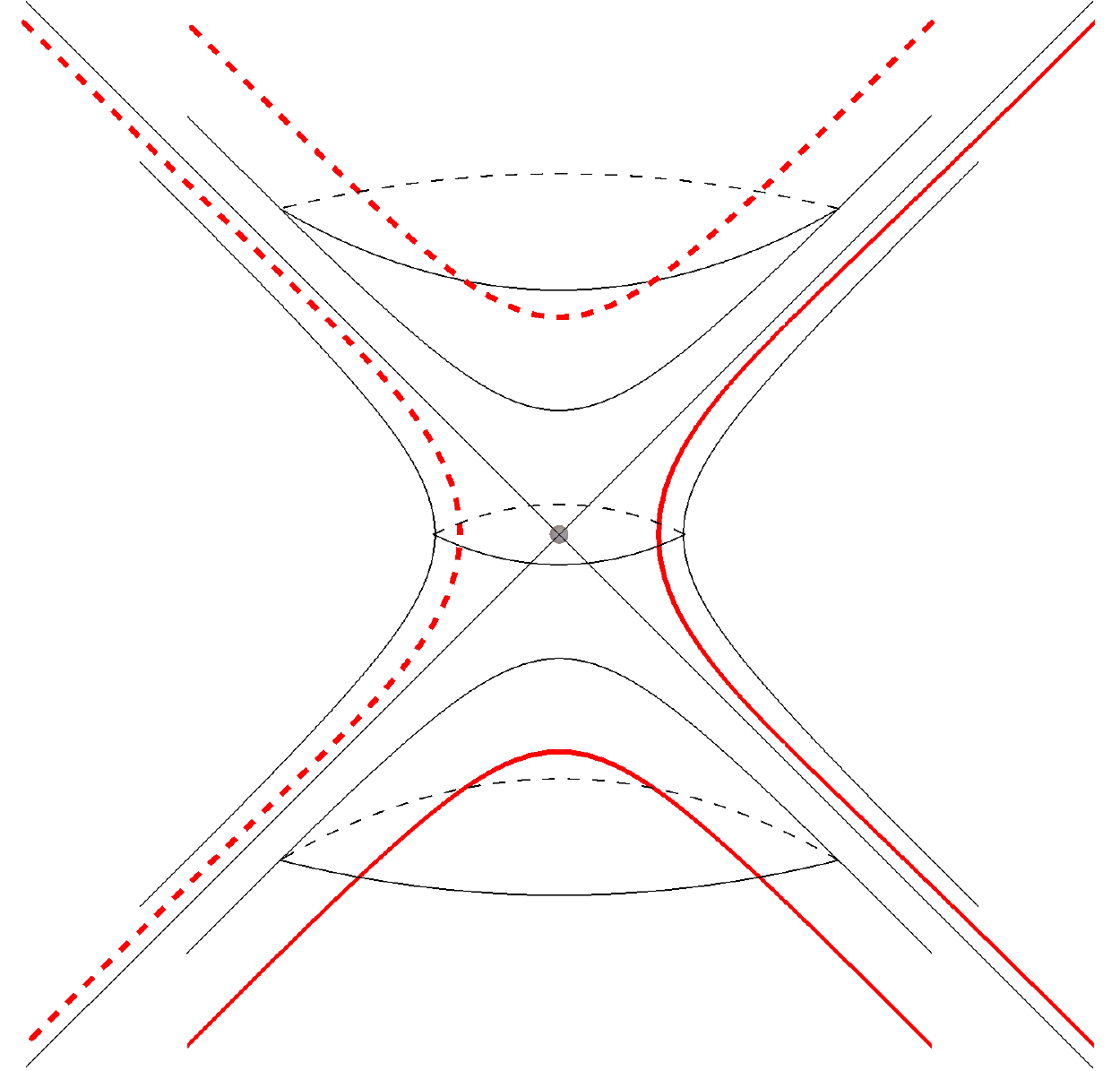}
\end{center}
\caption{A Hyperbolic Action on the Orbits of $sl(2,\mathbb{R})$.}
\label{HyperbolicActionOnSL2ROrbits}
\end{figure}

\subsection{A Hyperbolic Action on Discrete Orbits}
\label{hypactionondiscrete}

 We wish to describe the action of a hyperbolic generator on the continuous and discrete orbits and define a coordinate that parameterizes the orbits of a hyperbolic one-parameter subgroup. 
 We start out with the discrete orbits.
A  parameterization of the discrete orbits adapted to the hyperbolic one-parameter subgroup generated by $Y_2$ is given by:
\begin{equation}
(Y_0,Y_1,Y_2) = \alpha (\pm \cosh \rho \cosh v,  \cosh \rho \sinh v, \sinh \rho) \, .
\label{DiscreteOrbitRhoAndV}
\end{equation}
The coordinates $(\rho,v)$ take values in the real line. 
The  sign choice corresponds to a choice of  sheet. The coordinate $v$ parameterizes the orbits of the action of the one-parameter subgroup. 
Note that one coordinate system suffices to cover the whole of the upper sheet, namely the discrete $D^+$ orbit.
We also introduce an alternative coordinate system that will be used later on. We  set $y=e^v$ and  parameterize the orbits in terms of $(y,Y_2)$:
\begin{equation}
(Y_0,Y_1) = \frac{\sqrt{(Y_2)^2+\alpha^2}}{2} \,\left(\pm (y+y^{-1}), (y-y^{-1}) \right) \, .
\end{equation}
The discrete plus orbit is covered by the range $y>0$. The range of the momentum $Y_2$ is the whole of the real line. The symplectic form in the $(y, Y_2)$ coordinates is $\Omega = \alpha\, dY_2\wedge d\log y$.

\subsection{The Continuous Orbits as Patchworks}
The parameterizations of the continuous orbit in terms of coordinates along the action of hyperbolic one-parameter subgroups  introduces various global subtleties. We discuss this in multiple coordinate systems -- all of them will play in a role in the following.

\begin{figure}
\begin{center}
\includegraphics[scale=0.5]{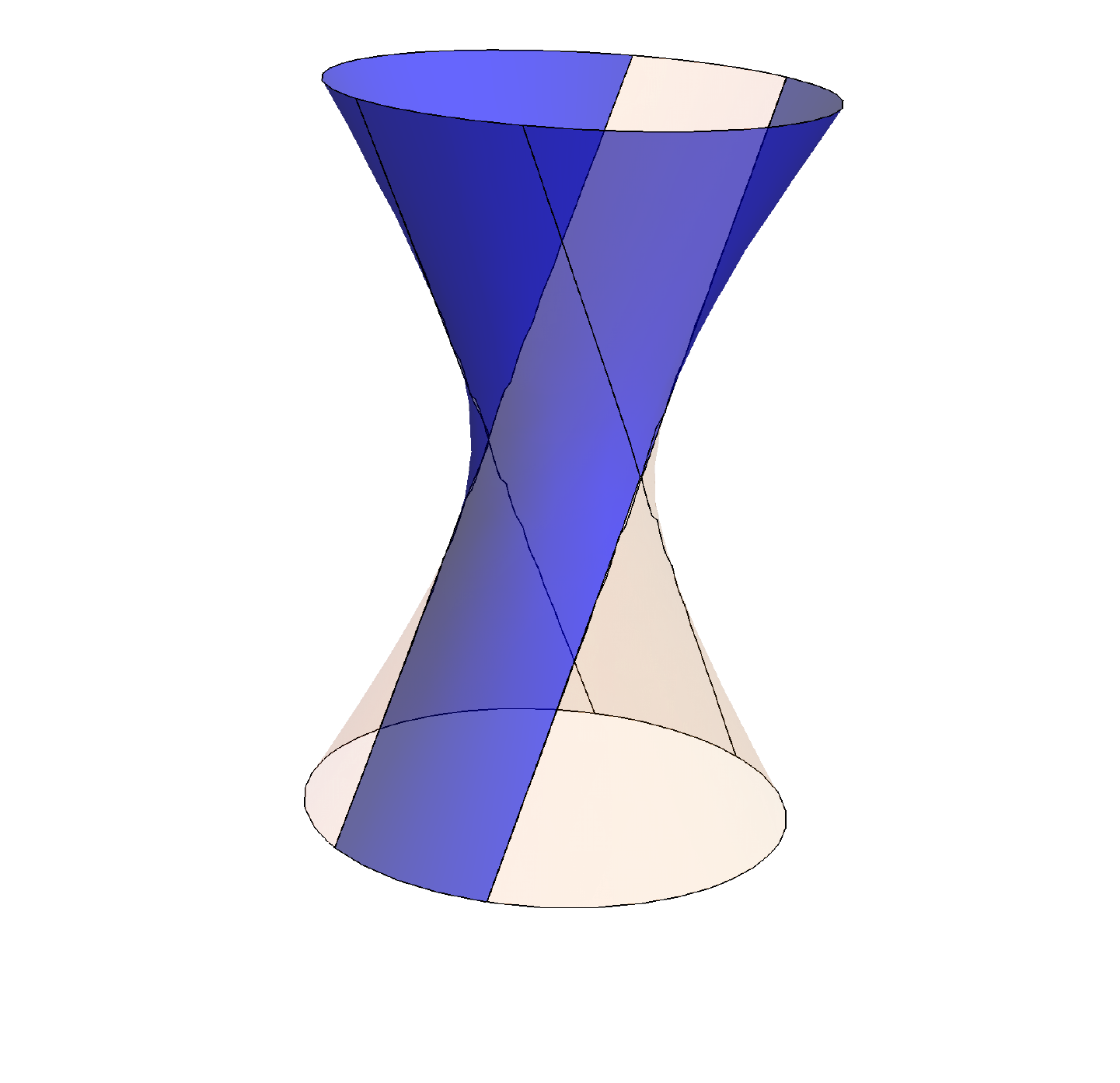}
\end{center}
\caption{Two regions of a continuous $sl(2,\mathbb{R})$ orbit.}
\label{OpenFunnels}
\end{figure}

Suppose we fix a point $Y_2$. Then the solution set of the orbit equation 
\begin{equation}
-(Y_0)^2 + (Y_1)^2 + (Y_2)^2 = \alpha^2 \,,
\end{equation}
corresponds to two (in general) disconnected curves forming a one-dimensional hyperbola. 
The boost in the $1$ direction connects points along each of these individual curves. Thus, we see that the hyperbolic action naturally leads to a two-sheeted structure. The first explicit parameterization to illustrate this generic feature is a $(v,Y_2)$ coordinate system. If the sign of $\alpha^2-(Y_2)^2$ is positive, we can parameterize the two branches as 
\be
\left(\sqrt{\alpha^2-(Y_2)^2}\, \sinh v,\pm \sqrt{\alpha^2-(Y_2)^2}\, \cosh v, Y_2 \right) 
\ee
 while if it is negative as  
 \be
 \left(\pm \sqrt{-\alpha^2+(Y_2)^2}\, \cosh v, \sqrt{-\alpha^2+(Y_2)^2}\, \sinh v, Y_2 \right) \, .
 \ee
 A global parameterization of the orbit is thus provided by the coordinate system $(v,Y_2)$, up to the change of branch on the one-dimensional hyperbola which is captured by the $\pm$ in front of the square root.  
 Note that if we use the coordinate system $(v,Y_2)$ and pick the upper branch in each region of $Y_2$ then we parameterize half of the orbit. For $|Y_2| < |\alpha|$ we parameterize the part with $Y_1>0$ and for $|Y_2| > |\alpha|$ the part with $Y_0>0$.
 Figure \ref{OpenFunnels} illustrates the two patches.   The patches join to make the  hour-glass orbit.
 The symplectic form is canonical in $(v,Y_2)$ variables. 

\begin{figure}
\begin{center}
\includegraphics[scale=0.5]{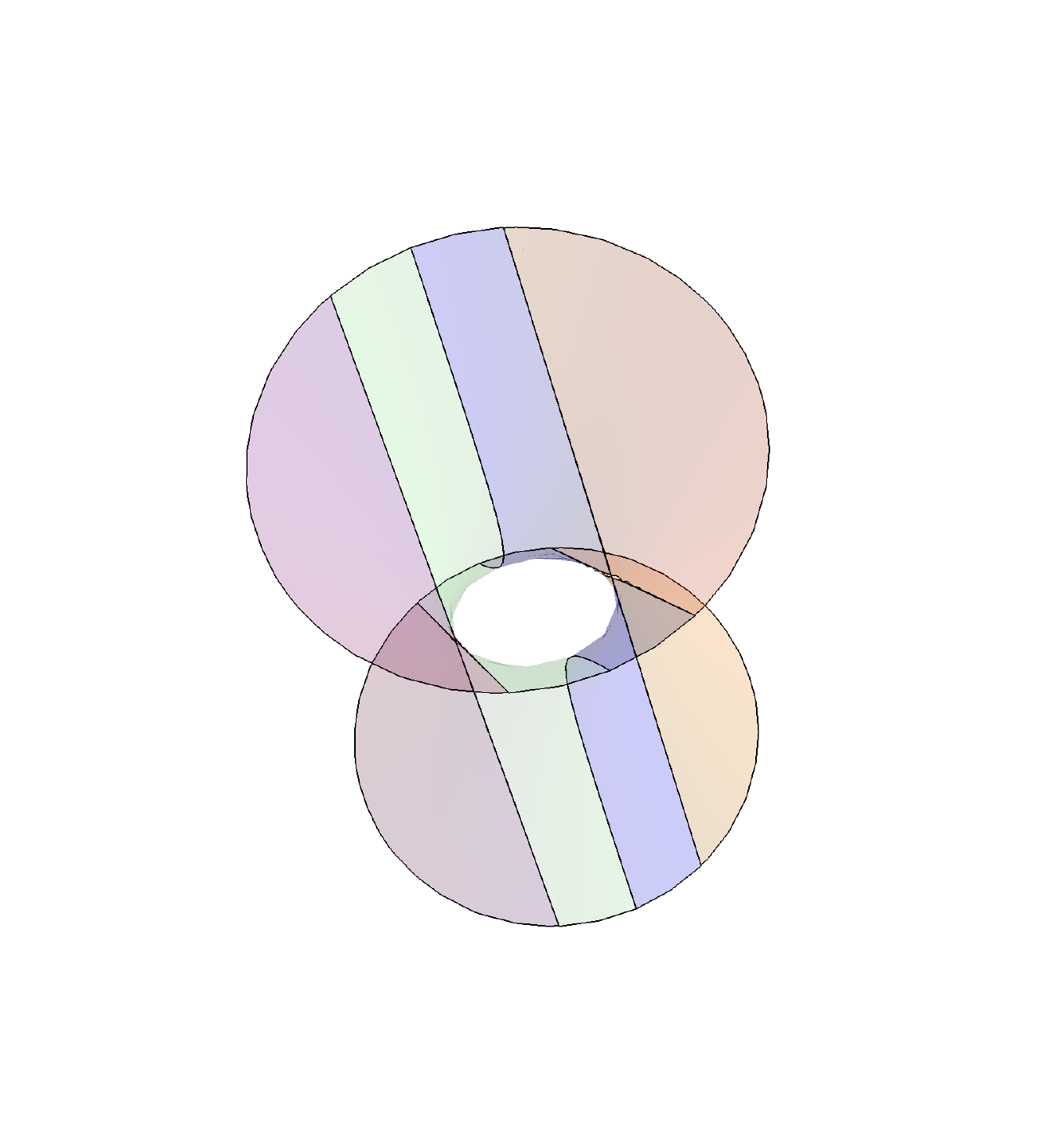}
\end{center}
\caption{Four regions of a continuous $sl(2,\mathbb{R})$ orbit.}
\label{HyperbolicStrips}
\end{figure}

A second manner to parameterize the orbit for $Y_2 \ge |\alpha|$ is:
\begin{equation}
(Y_0,Y_1,Y_2)= \alpha\, \left(  \sinh \rho \cosh v,  \sinh \rho \sinh v, \pm \cosh \rho \right) \, .
\label{RoevOne}
\end{equation}
%
For $|Y_2| \le |\alpha|$, we can parameterize the points on the orbit as:
\begin{equation}
(Y_0,Y_1,Y_2)= \alpha\, \left(  \sin \rho \sinh v,  \sin \rho \cosh v,  \pm \cos \rho \right) \, .
\label{RoevTwo}
\end{equation}
We again see the two-sheeted structure. Each $(v,\rho)$ patch (where $\rho \in ] -\pi/2,\pi/2[$ in the trigonometric case) covers a $Y_2$ patch of fixed (positive or negative) sign. We  illustrate the patches in Figure \ref{HyperbolicStrips}. 
The hour-glass is cut into three regions, and the middle region can be separated in two more according to the sign of $Y_2$. The four patches correspond then to the four choices of signs available above. 

Finally, a third parameterization is useful. It has the feature of parameterizing the two branches of the $(Y_0,Y_1)$ hyperbola at once. We define the points $(x,Y_2)$ solving the orbit equation for $|Y_2|\le \alpha$:
\begin{equation}
\label{Y2<alphapatch}
\left( \frac12(x-\frac{1}{x}) \sqrt{\alpha^2-(Y_2)^2}, \, \frac12  (x+\frac{1}{x}) \sqrt{\alpha^2-(Y_2)^2},\, Y_2 \right) \, ,
\end{equation}
as well as, for $|Y_2| \ge \alpha$:
\begin{equation}
\label{Y2>alphapatch}
\left( \frac12 (x+\frac{1}{x}) \sqrt{(Y_2)^2-\alpha^2},\,   \frac12 (x-\frac{1}{x}) \sqrt{(Y_2)^2-\alpha^2},\, Y_2\right) \, .
\end{equation}
The two branches correspond to $x>0$ and $x<0$ respectively. The $(x,Y_2)$ coordinate system covers the continuous orbit in a satisfactory global manner. The relation to the other two coordinate systems, whenever their regimes of validity overlap, is straightforwardly computed. 
In all these coordinates and patches, we can easily compute the volume form and an action appropriate to any given patch. We skip the detailed descriptions. One basic result to note is that the symplectic form in the $(x,Y_2)$ coordinate system is  $ \Omega=\alpha \, d Y_2 \wedge  d \log x$.

\subsubsection{The  Continuous Orbit and the Group }
\label{OrbitAndGroup}
It is rewarding to consider the relation between a parameterization of the SL$(2,\mathbb{R})$ group manifold and the continuous orbit in greater detail. This exercise helps further in visualizing aspects of our analysis and clarifies the global geometry. 
Very concretely, we can consider an explicit two-dimensional representation of the real Lie algebra $sl(2,\mathbb{R})$ in terms of Pauli matrices:
\begin{equation}
\sigma_0 = \begin{pmatrix}
0 & 1\\
-1 & 0
\end{pmatrix} = i \sigma_2
\qquad \quad
\sigma_1 = 
\begin{pmatrix}
0 & 1\\
1 & 0
\end{pmatrix}
\qquad \quad
\sigma_3 = \begin{pmatrix}
1 & 0\\
0 & -1
\end{pmatrix} \, .
\end{equation}
The coordinates on our Lie algebra $Y_\mu$ can be chosen  such that a Lie algebra element $t$ equals $t=Y_0 \sigma_0 + Y_1 \sigma_1 + Y_2 \sigma_3$. The Killing norm of the Lie algebra element $t$ is then  proportional 
to $-(Y_0)^2+(Y_1)^2+(Y_2)^2$ as before. 
Moreover, we can exponentiate this standard parameterization into the SL$(2,\mathbb{R})$ group manifold. In the group SL$(2,\mathbb{R})$, almost every element  can be factorized as:
\be
g = d_1 (-e)^{\epsilon_1} s^{\epsilon_2} p \, d_2~,
\ee
where $\epsilon_{i} = 0$ or $1$, the central group element $(-e)$ is minus the identity matrix, the matrices $d_1,d_2$ span hyperbolic one-parameter subgroups and the matrix $s$ squares to minus one:  
\begin{align}
-e &= 
\begin{pmatrix}
-1 & 0 \\
0 & -1
\end{pmatrix}
~,\qquad\hspace{1.3cm} 
s =
\begin{pmatrix}
0 & 1\\
-1 & 0
\end{pmatrix}
\nonumber \\
d_1 &= 
\begin{pmatrix}
e^{v/2}  & 0  \\
0 & e^{-v/2}
\end{pmatrix}
~, \qquad \quad
d_2 =
\begin{pmatrix}
e^{\psi/2} & 0\\ 
0 & e^{-\psi/2}
\end{pmatrix} \, .
\end{align}
The matrix $p$ on the other hand can take a hyperbolic or a trigonometric form:
\begin{align}
p_{h} &= 
\begin{pmatrix}
\cosh \rho/2 & \sinh \rho/2 \\
\sinh \rho/2 & \cosh \rho/2 
\end{pmatrix}, \hspace{.75cm} 
p_t = 
\begin{pmatrix}
\cos \rho/2 & \sin \rho/2\\
-\sin \rho/2 & \cos \rho/2 
\end{pmatrix}\, .
\end{align}
The parameter $\rho$ runs over $]-\infty, +\infty[$ for the hyperbolic group element $p_h$ and over the interval $]-\pi/2,\pi/2[$ for elliptic group elements $p_t$. This parameterization of the group is  adapted to the parameterization of continuous orbits of the Lie algebra. To see this in detail, it is sufficient to consider the orbit of a Lie algebra element $t=\alpha \sigma_3$. When we conjugate this element with a group element $g$ in the parameterization above, we find that the resulting Lie algebra element does not depend on the coordinate $\psi$. We obtain a parameterization of the orbit in terms of the coordinates $(v,\rho)$. The group element $(-e)=(-1)$ is central in SL$(2,\mathbb{R})$ and  does not act on the  PSL$(2,\mathbb{R})/\mathrm{SO}(1,1)$ coset. We can therefore distinguish four patches in the orbit, associated to a choice of $p_{h}$ or $p_t$ and whether the group element $s$ is present in the group element $g$ or not. The conjugated Lie algebra element has a $Y_2$ coordinate equal to  $\pm \cosh \rho$ for $p_h$, where the sign depends  on whether the group element $s$ is present or not, and $\pm \cos \rho$ for the choice of trigonometric $p_t$.
This coordinate system agrees with the second parameterization above in equations \eqref{RoevOne} and \eqref{RoevTwo}.
 The coordinate $v$ still parameterizes the hyperbolic action (from the left) on the PSL$(2,\mathbb{R})/\mathrm{SO}(1,1)$ coset and indeed coincides with the coordinate $v$ introduced previously. Similarly, the $Y_2$ coordinate is identified with $\pm \cosh \rho$ or $\pm \cos \rho$ up to rescaling by $\alpha$. The insertion of $s$ changes the sign of the $Y_2$ coordinate, namely of the momentum conjugate to the coordinate $v$. 

We note that as a consequence of our analysis, the continuous orbit provides a rather faithful two-dimensional representation of the three-dimensional group PSL$(2,\mathbb{R})$, provided we are prepared to ignore one topologically trivial hyperbolic direction. In particular, note that the first homotopy group of the group equals $\Pi_1(\text{PSL}(2,\mathbb{R}))=\mathbb{Z}$ as recalled in the introductory subsection \ref{Topology}. The continuous orbit indeed exhibits the topologically non-trivial circle faithfully.

\section{The Path Integral on the Discrete Orbits}
\label{DiscreteOrbits}
In this section, we perform the phase space path integral over the discrete orbit, in coordinates that are chosen to render a one-parameter hyperbolic action simple.  We start out with a canonical analysis of the system and its symmetries. We then integrate over discretized paths. Because the discrete orbit can be covered with a single patch, it is a good warm-up example for the continuous orbit in which we need at least two patches.

\subsection{The Classical System and its Symmetries}
The symplectic form $\Omega$ on the discrete $D^+$ orbit equals 
\begin{equation}
\Omega =  \alpha \cosh \rho d \rho \wedge dv = \alpha d(\sinh\rho\, dv)~.
\ee
The primitive one-form pulled back to the world line of a particle defines the Lagrangian ${\cal L}$ of our quantum mechanics:
\be
{\cal L} = \alpha \sinh \rho \, \dot{v} + \beta\, \dot{v}~.
\label{LagrangianDiscretePlusOrbit}
\ee
We have added a total derivative term.
The  momentum $\pi_v$ conjugate to the coordinate $v$ equals:
\be
\pi_v = \frac{\pa {\cal L}}{\pa \dot{v}} = \alpha\sinh\rho +\beta ~.
\ee
The total derivative term allows us to shift the momentum by a constant $\beta$.
We can write the $sl(2,\mathbb{R})$ symmetry generators $Y_i$ as functions on phase space:
\begin{align}
Y_2 &= \alpha \sinh \rho = \pi_v-\beta
 \\
Y_0 &= 
\alpha \cosh v \cosh \rho =  \cosh v \sqrt{\alpha^2 + (\pi_v-\beta)^2}
 \\
Y_1 &= \alpha \sinh v \cosh \rho = \sinh v \sqrt{\alpha^2+(\pi_v-\beta)^2}~.
\end{align}
If we think of the orbit as the right geometric coset PSL$(2,\mathbb{R})/SO(2)$, then these operators
generate a left $sl(2,\mathbb{R})$ action on the orbit.
The   brackets between functions on phase space equals
\be
\{ f(v, \pi_v), g(v,\pi_v)\} = \frac{\pa f}{\pa v} \frac{\pa g}{\pa \pi_v} - \frac{\pa g}{\pa v} \frac{\pa f}{\pa \pi_v}  ~.
\ee
The Poisson brackets between the $sl(2,\mathbb{R})$ generators $Y_i$ are:
\begin{align}
\label{PBys}
\{ Y_0,Y_1 \} = -
 Y_2~, \qquad
\{ Y_1, Y_2 \} = 
Y_0~,\qquad
\{ Y_2, Y_0 \} = 
-Y_1~.
\end{align}
The transitive group action on the phase space will translate into a representation space in the quantum theory which is an irreducible representation of the group. The classical symmetry generators will turn into quantum operators satisfying a $sl(2,\mathbb{R})$ algebra.

\subsubsection{Towards the Quantum Theory}

Keeping in mind the quantum mechanical analysis to follow, we define generators $Y_\pm= (Y_0 \pm Y_1)$ and factorize:
\be
Y_2= \pi_v - \beta
~,\qquad 
Y_{\pm}= e^{\pm v} \sqrt{(\pi_v-\beta+i \alpha)(\pi_v-\beta - i \alpha)  }~.
\ee
We can   rescale the generators $Y_{\pm}$ by purely momentum dependent factors as this leaves invariant the Poisson brackets  \eqref{PBys}. We thus eliminate square roots:
\begin{align}
Y_2  &=  \pi_v -\beta
\nonumber \\
Y_+ &= e^v ( \pi_v -\beta +i \alpha) \nonumber \\
Y_- &= e^{-v} (\pi_v -\beta-i\alpha)\ .
\end{align}
From now on, we will parameterize the coefficient of the action in terms of a shifted parameter ${j}$ as  $\alpha={-j}+1/2$.\footnote{See footnote \ref{VolumeFootnote} for an intuitive understanding of the shift of the spin $j$ by $1/2$ in the case of finite dimensional representations. In the non-compact context, one can consider the geometric picture of where discrete and continuous orbits meet, at $j=1/2$, to justify this shift pictorially.} Moreover, we will pick our origin of momentum by setting  $i\beta=- \, j =\alpha-1/2$.
  Moreover, it is important to note that the generators $Y_{\pm}$ contain a normal ordering ambiguity proportional to $e^{\pm v}$ in the quantum theory.
We can tune the normal ordering constant such that 
our final expressions for the quantum operators $Y_i$  take the form
\begin{align}
\label{rescaledys}
Y_2  &=  \pi_v - i j  \nonumber \\
Y_+ &= e^v ( \pi_v -2i{j} ) \nonumber \\
Y_- &= e^{-v} \pi_v \ .
\end{align}

\subsection{Bases of States} 
\label{discretewavefunctions}

Before discussing the path integral let us first describe in a little bit of detail the Hilbert space of wavefunctions on which the differential operators $Y_a$ act. 
We shall realize the discrete series representations on functions on the  half-line, parameterized by $y$. 
We shall identify the coordinate of the half-line with the exponential of the coordinate $v$ of the discrete orbit: 
\be 
y = e^v~.
\ee
In terms of this coordinate, the $sl(2,\mathbb{R})$ generators take the form  -- we use $\pi_v = -i\pa_v = -iy \pa_y$ --:
\begin{align}  
Y_2 &= -iy\pa_y - i \, {j} \cr
Y_+ &= -iy^2 \pa_y - 2i  {j} y \cr
Y_- &= -i\pa_y ~.
\label{ysonfy}
\end{align}
We work in a setting in which we diagonalize the hyperbolic generator $Y_2$, which  amounts to diagonalizing the momentum operator. Given the form of the differential operator above, the momentum eigenstates can be written as:
\be 
\label{wavefunoniy}
\zeta_{\mu}(y) 
= 
y^{-\mu}~. 
\ee 
%
%
The $Y_2$ eigenvalue of the wavefunction $\zeta_{\mu}(y)$ can be  calculated using the differential operator \eqref{ysonfy}, and is given by $i(\mu{-j})$. Implicitly, we define the momentum wave-function $\Phi$ as a Mellin transform of the position wave-function $f$:
\begin{equation}
\Phi(\lambda) = \int_0^\infty~ dy f(iy)~ y^{\lambda-1} \, ,
\end{equation}
where we analytically continue the wave-function $f$ from the positive imaginary axis $iy$ to the whole of the upper half plane $z=iy$. The inverse transform is given by:
\begin{equation}
f(iy) = \frac{1}{2 \pi i} \int_{a-i \infty}^{a+i \infty} d\mu~\Phi(\mu) y^{-\mu } \, .
\end{equation}
The integral over the eigenvalues is along 
 a shifted imaginary axis:
\be 
\mu \in {{a}} + i\, \mathbb{R} \, ,
\ee 
where the shift $a$ is chosen such that the integral is well-defined. 
The  wavefunctions $\zeta_{\mu}$  satisfy the completeness relation:
\be 
\label{ynorm}
\int_0^{\infty}\frac{dy}{y}~ \zeta_{\lambda}^{\star}(y)~\zeta_{\mu}(y) = \int_0^{\infty}\frac{dy}{y}~y^{\lambda-\mu} = 2\pi i\, \delta(\lambda-\mu)~.
\ee 
The last equality follows when $\lambda-\mu$ is purely imaginary. Our integration measure is dictated by the symplectic form on the discrete orbit. We shall see this more explicitly when we discuss the path integral formulation.
We now define normalized kets $|\mu\rangle$ such that their wavefunctions defined on the half-line are given by $\zeta_{\mu}(y)$ in \eqref{wavefunoniy}.
These satisfy the completeness relation in momentum space:
\begin{align}
    %
    \frac{1}{2\pi i }\int_{a-i\infty}^{a+i\infty}~d\mu ~ |\mu\rangle \langle \mu| =1 ~. 
\end{align}

\subsection{The Path Integral}

Given the action of the generators  \eqref{ysonfy} on the space of functions on the half-line it is possible to proceed via the Hamiltonian formalism and obtain the matrix elements for these and the exponentiated operators, leading to the well known integral expressions for the kernels that exhibit  the PSL$(2,\mathbb{R})$ group action on the discrete representation labelled by real and integer values of the parameter $j$ \cite{PinkBook}. Our goal in the present section is to derive these from a path integral calculation on the orbit corresponding to the discrete representation. 

We shall work with the phase space path integral. As we saw in Section \ref{hypactionondiscrete}, there are different ways to parameterize the orbit. A particularly convenient choice is to use $(y, \eta)$ variables, where $y=e^v$ and $\eta = \sinh\rho$. From the classical expressions in \eqref{rescaledys} we see that $\eta$ is related to the canonically conjugate momentum by the relation 
\be 
\alpha\, \eta = \pi_v - i \, {j} ~.
\ee 
We consider a discretized version of the path integral, in which the total time $T$ between initial and final states is split into $N$ time intervals of width $\Delta t$. We perform the path integral specifying the initial and final momenta. In this case, the discretized path integral integrates over $N-1$ momenta $\eta_a$ and $N$ intermediate positions $y_a$. The action \eqref{LagrangianDiscretePlusOrbit} that weights the path integral is obtained from the symplectic structure on the orbit. Since we fix the initial and final momenta it turns out to be more useful to add a total derivative term to that action and begin with the  discretized action:
\be 
S =  -\alpha\int~ dt~ \log y(t)\, \dot{\eta}(t) = -\alpha\sum_{a=1}^N \log y_a(\eta_a - \eta_{a-1}) ~.  
\ee 
The initial and final $\eta$ values are given in terms of the initial and final momenta $(i\,\mu, i\, \lambda)$:
\be 
\label{initialfinalmom}
\eta_0 = \frac{i}{\alpha}(\mu - {{j}})~, \quad \eta_N = \frac{i}{\alpha}(\lambda - {{j}})~. 
\ee 
 Thus the matrix element for an arbitrary operator ${\cal O}$ is given by the  discretized path integral:
\begin{equation}
\langle \lambda | {\cal O} | \mu\rangle = \int_0^{\infty}\prod_{a=1}^{N} \frac{dy_a}{y_a} \int_{-\infty}^{\infty} \prod_{a=1}^{N-1}\frac{d \eta_a}{\alpha}~ e^{iS}~ {\cal O} ~. 
\end{equation}

\subsubsection{Kernel for the Hyperbolic Generator}

Our first goal is to compute the momentum space kernel that represents the group action in the discrete representation labelled by $j$:
\be 
K(\lambda, \mu; {{j}}, e^{i T y_2}) =\frac{1}{2\pi i} \langle \lambda | e^{i T Y_2} | \mu\rangle~. 
\ee 
We recall that in terms of the phase space coordinates the symmetry generator $Y_2$ equals $Y_2 = \alpha \eta$. This particular matrix element can therefore be calculated by modifying the discretized action in the following manner:
\begin{align}  
S_{hyp} &= -\alpha\sum_{a=1}^N \log y_a(\eta_a - \eta_{a-1})  + \sum_{a=1}^{N-1} \Delta t\, \alpha\, \eta_a \cr
&= \alpha \eta_0 \log y_1 - \alpha \eta_N \log y_N + \sum_{a=1}^{N-1} \alpha\eta_a(\log y_{a+1} - \log y_a + \Delta t)~, 
\end{align}
where $\eta_0$ and $\eta_N$ are given in equations \eqref{initialfinalmom}. Substituting this into the path integral we see that all the integrals over the $N-1$ variables $\eta_a$ can be trivially done, and this sets $\log y_{a} = \log y_{a+1} + \Delta t$. This eliminates all but one of the $y_a$-integrals and using $\log y_1 = \log y_N + T$, we obtain 
\begin{align}
    K(\lambda, \mu; {{j}}, e^{i T Y_2}) &=\frac{1}{2\pi i}  \int_0^{\infty}\frac{dy}{y} e^{i\alpha(\eta_0 - \eta_N)\log y }\, e^{i\alpha\eta_0 T}\cr
    &= \frac{1}{2\pi i} \int_0^{\infty}~\frac{dy}{y}~ y^{\lambda-\mu}~e^{-(\mu-{{j}})T } =  \delta(\lambda-\mu)\, e^{-(\mu-{{j}}) T } ~.
\end{align}

\subsubsection{Kernel for the Parabolic Generators}

The derivation of the kernel for the parabolic elements is a little more involved. We first compute the kernel for an infinitesimal time interval, which is equivalent to calculating the  matrix element:
\be 
\langle\lambda|Y_-|\mu\rangle~. 
\ee 
The operator $Y_-$ can be written down in terms of the phase space variables as 
\be 
Y_- = y^{-1}(\alpha\, \eta +i\, {{j}} )~. 
\ee 
Here we have to choose a convention for which $\eta$-value to insert, namely whether it is the initial one in the infinitesimal interval, the final one, or some combination. We choose the initial value of the interval in what follows. One thus has a single interval, with action given by 
\begin{align}
    S = \alpha(\eta_0-\eta_1) \log y = i(\lambda-\mu)\log y~,
\end{align}
and a single $y$-integral to do:
\begin{align}
    \langle\lambda|Y_-|\mu\rangle &= \int_0^{\infty}\frac{dy}{y}~y^{-1} (\alpha\eta_0 +i {{j}})~y^{\lambda-\mu}  \cr
    &= (i\, \mu)\, 2\pi i\, \delta(\lambda - \mu - 1)~.
\end{align}
Given this matrix element, we can  derive the matrix element for the exponentiated $Y_-$ operator in the representation labelled by $ {{j}}$ by repeated insertion of the complete set of momentum eigenstates. For $c>0$ we have 
\begin{align}
K(\lambda, \mu;  {{j}}, e^{c Y_-}) &=\frac{1}{2\pi i} \langle \lambda| e^{c Y_-}|\mu\rangle \cr
&=\frac{1}{2\pi i} \sum_{n=0}^{\infty} \frac{c^n}{n!}\int  \frac{d\mu_1}{2\pi i}\ldots \int \frac{d\mu_n}{2\pi i} \langle\lambda|  Y_-| \mu_1\rangle \prod_{i=1}^{n-1} \langle\mu_i|Y_-| \mu_{i+1}\rangle
\langle\mu_n| Y_-| \mu\rangle\cr
&=\frac{1}{2\pi i}\sum_{n=0}^{\infty}(i c)^n \frac{\mu(\mu+1)\ldots (\mu+n-1)}{n!} (2\pi i)\delta(\lambda-\mu-n) \cr
&=\frac{1}{2\pi i}\int_{0}^{\infty}~ \frac{dy}{y}~ \sum_{n=0}^{\infty}(i c)^n \frac{\mu(\mu+1)\ldots (\mu+n-1)}{n!} y^{(\lambda-\mu - n)}\cr
&= \frac{1}{2\pi i}\int_{0}^{\infty}~ \frac{dy}{y}~ y^{(\lambda-\mu)} (1-i y^{-1}c)^{-\mu}\cr
&=\frac{1}{2\pi i} \int_0^{\infty}~dy~y^{\lambda-1}(y-ic)^{-\mu}
~. 
\end{align}
One can repeat the analysis for the $Y_+$ generator. In terms of the phase space variables we have
\be 
Y_+ = y (\alpha \eta - i \, {{j}}) ~.
\ee 
Performing the path integral for the infinitesimal interval we obtain 
\begin{align}
    \langle\lambda|Y_+|\mu\rangle &= \int_0^{\infty}\frac{dy}{y}~y (\alpha\eta_0 -i  {{j}})~y^{\lambda-\mu}  \cr
    &= i\, (\mu-2 j)\, 2\pi i\, \delta(\lambda - \mu + 1)~.
\end{align}
One can now compute the kernel for the $Y_+$ operator as before. For $b>0$ we have
\begin{align}
K(\lambda, \mu;  {{j}}, e^{b Y_+}) &=\frac{1}{2\pi i} \langle \lambda| e^{b Y_+}|\mu\rangle \cr
&=\frac{1}{2\pi i} \sum_{n=0}^{\infty} \frac{b^n}{n!}\int \prod_{i=1}^n \frac{d\mu_1}{2\pi i}\ldots \int \prod_{i=1}^n \frac{d\mu_n}{2\pi i} \langle\lambda| Y_+| \mu_1\rangle \prod_{i=1}^{n-1} \langle\mu_i|Y_+| \mu_{i+1}\rangle
\langle\mu_n| Y_+| \mu\rangle\cr
&=\frac{1}{2\pi i}\sum_{n=0}^{\infty}(i b)^n \frac{(\mu-2 {{j}})(\mu-2 {{j}} -1)\ldots (\mu-2 {{j}} -n+1)}{n!} (2\pi i)\delta(\lambda-\mu+n) \cr
&=\frac{1}{2\pi i}\int_{0}^{\infty}~ \frac{dy}{y}~ \sum_{n=0}^{\infty}(i b)^n \frac{(\mu-2 {{j}})(\mu-2 {{j}} -1)\ldots (\mu-2 {{j}} -n+1)}{n!} y^{(\lambda-\mu + n)}\cr
&= \frac{1}{2\pi i}\int_{0}^{\infty}~ dy~ y^{\lambda-1}~ y^{-\mu}~ (1+i\, b\, y )^{\mu-2 {{j}}} 
~.
\end{align}
Thus, we have computed the action of the exponentiated infinitesimal generators $Y_{2}$ and $Y_{\pm}$. 

\subsubsection{The Insertion s} 
\label{FirstDiscussionOfMinusE}
As will be manifest shortly, it is useful to consider the matrix element of the $s$-operator between momentum eigenstates. As a two-by-two matrix in the fundamental representation of PSL$(2,\mathbb{R})$ the operator $s$ equals the matrix $\sigma_0 = i\sigma_2$. When we conjugate a general Lie algebra element $t=Y_0\sigma_0 + Y_1\sigma_1+Y_2\sigma_3$ by this group element, we see that it acts as $(Y_0,Y_1, Y_2) \longrightarrow (Y_0,-Y_1, -Y_2)$. On the phase space variables, the $s$-operator therefore flips the sign of $\eta$ and  $y$.\footnote{See e.g. equation \eqref{DiscreteOrbitRhoAndV} and use the relation  $\eta=\sinh \rho$.}
Moreover, the sign flip on the label of the momentum states can be accompanied by an extra phase factor action in the quantum theory. To determine the possibilities for this phase, it is
 important to note that the relation $s^2= (-e)$ must hold in the group SL$(2,\mathbb{R})$. Thus, the phase squared must be $\pm 1$.
 
 We can gather further information about this phase from the elliptic point of view provided in Section \ref{ellipticquant}. Firstly, we note from equation \eqref{DiscreteOrbitRAndPhi} that the $s$ operator acts to shift the elliptic coordinate $\phi \rightarrow \phi-\pi$. Thus, when we insert an $s$-operator in the path integral, we pick up a sign $e^{-i \pi \epsilon}=e^{-i \pi j}$ from the second term in the action \eqref{DiscreteOrbitEllipticAction}. Thus, we associate this extra phase to the action of the operator $s$. For discrete representations of PSL$(2,\mathbb{R})$ the index $j$ is integer. The operator $s$ then squares to the identity. For representations of SL$(2,\mathbb{R})$ the parameter $j$ can be integral or half integral and minus the identity $(-e)$ inserts a phase $(-1)^{2j}$.\footnote{We revisit these points in more generality when we discuss the path integral realization of orbits of SL$(2,\mathbb{R})$ and its universal cover in Section \ref{repsofcovers}. }
With all this in mind the path integral that defines the matrix element for the $s$ operator in the discrete representation labelled by $j$ is given by
\begin{align}
    \langle \lambda| s |\mu\rangle =\frac{(-1)^{-j}}{2\pi i} \int_0^{\infty} \frac{dy}{y} ~ e^{i S'}~,
\end{align}
where the modified action $S'$ takes into account the fact that the initial momentum $\eta_0$ has to be sign-flipped:
\be 
S' = - (\alpha \eta_0 + \alpha \eta_1) \log y = -i(\mu+\lambda - 2{{j}})\log y~. 
\ee 
Finally, we conclude that the matrix element of the insertion $s$ equals
\begin{align}
    \langle \lambda| s |\mu\rangle 
    &= \frac{(-1)^{{{-j}}}}{2\pi i} \int_0^{\infty} \frac{dy}{y}~ y^{\lambda +\mu - 2{{j}}} = (-1)^{-j}\delta(\lambda +\mu - 2{{j}}) ~.
\end{align}
\subsubsection{Summary}

We have so far calculated the matrix elements of the hyperbolic and parabolic generators of PSL$(2,\mathbb{R})$ and of the $s$-operator. A key point is that  any group element $ g 
\in $ PSL$(2,\mathbb{R})$ can be decomposed in either of the following ways \cite{PinkBook}:
\begin{align}
\label{sl2elementdecomp}
    g &= g_-(t) \, \delta(x) ~, \cr
    g &= g_-(t_1)\, \delta(x) \, s \, g_-(t_2)~,
\end{align}
where 
\begin{align} 
g_-(t) = \begin{pmatrix} 1 &0 \\ t &1\end{pmatrix},\quad  \delta(x)= \begin{pmatrix} x &0 \\ 0 & x^{-1}\end{pmatrix}~,\quad s= \begin{pmatrix} 0 &1 \\ -1 &0\end{pmatrix}~. 
\end{align}
The first decomposition holds for the case when the right upper entry $b$ is zero, while the second holds when it is non-zero. Given that we have already obtained the kernels for all the factors  that appear in the decomposition, and by continuity in the group element, it follows that for a PSL$(2,\mathbb{R})$ group element of the form $g$, the kernel in momentum space reads
\begin{align}
\label{kerneldiscreteony}
K(\lambda,\mu;j,g) &= \frac{1}{2 \pi i} \int_0^\infty dy~ y^{\lambda-1}~ (a\, y -i\, c)^{-\mu}~ (i\, b\, y + d)^{\mu - 2{{j}}} \cr
&= \frac{1}{2 \pi i} \int_0^\infty \frac{dy}{y}~ y^{\lambda} \int_0^\infty dy'~ y'^{-\mu}\, (i\, b\, y+d)^{- 2{{j}}}~\delta\left(y'- \frac{ a\, y -i\, c}{i\, b\, y + d}\right)~ .
\end{align}
From the second equality one can read off the kernel in position space: 
\be
K(y,y'; j, g) =  (i\, b\, y+d)^{- 2{{j}}}~\delta\left(y'- \frac{ a\, y -i\, c}{i\, b\, y + d}\right)~.
\ee 
%
This concludes our path integral derivation of the kernel for the discrete representations of PSL$(2,\mathbb{R})$. The final result matches the literature.

 \section{The Path Integral on the Continuous Orbits}
 \label{HyperbolicPathIntegral}
 \label{ContinuousOrbits}

In this section we perform the path integral over continuous orbits. Unlike the case of the discrete orbits, one needs at least two patches to cover the orbit. That renders the path integration more subtle. We begin with a canonical analysis of the system in a  patch and derive expressions for the generators of the sl$(2,\mathbb{R})$ algebra. We then turn to the quantum theory and determine the irreducible representation to which the path integral over patches gives rise.
 
\subsection{The Canonical Classical Analysis}

In order to be concrete, we first work in the $Y_2 > \alpha$ patch,  in which we can set $Y_2 = \alpha \cosh\rho$. From equation \eqref{Y2>alphapatch} we have 
\begin{align} 
(Y_0, Y_1, Y_2) = \alpha\big( \frac12 (x+\frac{1}{x}) \sinh\rho,   \frac12 (x-\frac{1}{x}) \sinh\rho, \cosh\rho\big) \, .
\end{align}
We specify the symplectic form $\Omega$:
 \be
 \Omega = \alpha  \, d \cosh \rho \wedge d\log x~.
 \ee
The Lagrangian of a point particle is again given by the pullback of a primitive one-form $\Theta$, which is defined by $\Omega = d\Theta$. For our analysis, it is convenient to further work in the $x>0$ subregion and introduce the variable $v= \log x $, in terms of which we obtain the Lagrangian:
 \be
 {\cal L} = \alpha\, \cosh\rho\,  \dot{v} + \beta\, \dot{v}~.
 \ee
 As before we have added a total derivative term with an arbitrary coefficient $\beta$, which leads to the conjugate momentum: 
 \be 
 \pi_v = \alpha \, \cosh\rho + \beta
 \ee
 We then have the expressions for the sl$(2,\mathbb{R})$ generators as functions on phase space:
 \begin{align}
 Y_2 &= \alpha\, \cosh\rho  = \pi_v-\beta
 \nonumber \\
 Y_0 &= \cosh v \sqrt{(\pi_v-\beta)^2-\alpha^2}
 \nonumber \\
 Y_1 &= \sinh v \sqrt{(\pi_v-\beta)^2-\alpha^2}\, .
 \end{align}
 The Poisson brackets between these generators are given by
 \begin{align}
 \label{YPBsY2>alpha}
     \{Y_0, Y_1\} = -Y_2~,\qquad \{Y_1, Y_2\} = Y_0~,\qquad \{Y_2, Y_0\} = -Y_1~.
 \end{align} 
 Defining again $Y_\pm=Y_0\pm Y_1$, we find:
 \begin{align}
 Y_\pm &= e^{\pm v}  \sqrt{(\pi_v-\beta+\alpha)(\pi_v-\alpha-\beta)}
 \, .
 \end{align}
 We rescale by momentum dependent factors and eliminate the square roots:
 \begin{align}
 \label{YsintermsofxY2>alpha}
 Y_2 &=  \pi_v-\beta
 \nonumber \\
 Y_+&= -i e^{v}  (\pi_v-\alpha-\beta)
 \nonumber \\
 Y_- &= ie^{-v} (\pi_v-\beta+\alpha)
 \, .
 \end{align}
 We now make a choice of parameters $\beta = i \, j$ and $\alpha = i({j} -\frac12)$ which will be convenient in the quantum theory.
 For the irreducible continuous representation under consideration we have that ${j} = \frac12 - is$, with $s\in\mathbb{R}$. 
 Recalling that $v=\log x$ and up to a normal ordering constant, we obtain the differential operators that represent the sl$(2,\mathbb{R})$ generators:
 \begin{align}
 Y_2 &=  -i\, x\, \pa_x -i\, {j}~,
 \nonumber \\
 Y_+&= - x^2 \pa_x -2\, {j}\, x ~,
 \nonumber \\
 Y_- &= \pa_x~.
 \end{align}
Remaining in the $x>0$ patch, one can similarly work out the formulae in the $|Y_2|<\alpha$ region, in which one can set $Y_2= \alpha\cos\rho$. From equation \eqref{Y2<alphapatch} we see that
\begin{align}
(Y_0,Y_1,Y_2)=   \alpha \big( \frac12(x-\frac{1}{x})\sin\rho, \frac12  (x+\frac{1}{x}) \sin\rho, \cos\rho\big) ~,
\end{align}
The particle Lagrangian in this region of the orbit takes the form
\be
 {\cal L} = \alpha\, \cos\rho\,  \dot{v} + \beta\, \dot{v}~,
 \ee
 where we have again defined $v=\log x$. The sl$(2,\mathbb{R})$ generators in this region are 
\begin{align}
    Y_2 &= \alpha\cos\rho = \pi_v-\beta ~,\cr
    Y_0 &= \sinh v\sqrt{\alpha^2-(\pi_v-\beta)^2}~,\cr 
    Y_1 &= \cosh v\sqrt{\alpha^2-(\pi_v-\beta)^2}~.
\end{align}
These satisfy the same Poisson brackets  \eqref{YPBsY2>alpha}. One can rescale the sl$(2,\mathbb{R})$ generators to get rid of the square roots and substituting $\alpha$ and $\beta$ as before, we find exactly the same differential operators  \eqref{YsintermsofxY2>alpha}. This completes the canonical analysis in the $x>0$ region.

In the $x<0$ region of the orbit, one can repeat the analysis step by step and one obtains the same form of the differential operators for the sl$(2,\mathbb{R})$ generators, with the substitution $x\rightarrow (-x)$. In particular this means that the $Y_{\pm}$ operators pick up an extra sign while the $Y_2$ operator remains unchanged. This will have  consequences in our subsequent discussion.

\subsection{Bases of States}

We have seen that the orbit for the continuous representation is  composed of distinct regions. One needs at least two patches to cover the entire orbit, with distinct expressions for the sl$(2,\mathbb{R})$ symmetry generators.
The patch with $x>0$ and the patch with $x<0$ are described separately. 
This leads us to introduce pairs of functions $(x_+^\lambda, x_-^\lambda)$ which form a basis of generalized functions  depending on the coordinate $x$. These are defined such that for $\text{Re} \lambda > -1$ and an infinitely differentiable function $\varphi(x)$, we have \cite{PinkBook}:
\begin{align}
    (x_+^\lambda, \varphi(x)) = \int_0^{\infty} dx~x^{\lambda} \varphi(x)  ~, \quad (x_-^\lambda, \varphi(x)) = \int_{0}^{\infty}dx~ x^{\lambda} \varphi(-x) ~.
\end{align}
For a fixed function $\varphi$, the inner product depends analytically on the momentum label $\lambda$ and one can continue it to the domain $\text{Re}\lambda  \le -1$, avoiding the negative real integers\footnote{At these points, the generalized functions have poles. The residues at these poles evaluate to derivatives of $\delta$-functions.}.

We will be interested in computing the overlaps of momentum eigenstates using the path integral. The wavefunctions in position space are a pair of generalized functions, and  one introduces a pair of momentum eigenstates as well, related to the position space eigenstates via  the Mellin transform:
\begin{align}
   |\lambda, \pm\rangle &= \int_{-\infty}^{\infty} \frac{dx}{x}~ x_{\pm}^{-\lambda}|x\rangle = \int_{0}^{\infty} \frac{dx}{x}~ x^{-\lambda}|\pm x\rangle ~.
\end{align} 
We defined a state $|+ x \rangle$ which is non-zero for positive $x$ only. 
One reads off the wavefunctions that correspond to momentum eigenstates:
\be 
\langle x| \lambda, \pm\rangle = x_{\pm}^{-\lambda}~. 
\ee 
The corresponding relations for the bra-vector are
\begin{align}
   \langle \lambda, \pm | &= \int_{-\infty}^{\infty} \frac{dx}{x}~ x_{\pm}^{\lambda}\langle x| = \int_{0}^{\infty} \frac{dx}{x}~ x^{\lambda} \langle \pm x | ~.
\end{align}
Given these basis vectors, a  state $| f \rangle$ in the momentum basis is given by a pair of functions obtained via the overlaps with $|\lambda, \pm\rangle$, and are related to the function in position space via the Mellin transform:
\be
\label{MTf}
\tilde f_{\pm}(\lambda) = \langle \lambda, \pm | f \rangle = \int_0^{\infty} \frac{dx}{x}x^{\lambda} f(\pm x) ~.
\ee
%
The convergence of these integrals are dependent on the behaviour of $f(x)$ near infinity. For functions that represent the continuous representations, one has that $|f(x)|\sim |x|^{-2j}$ for large $x$ and  these integrals are absolutely convergent near infinity for $\text{Re}(\lambda -2j) < 0$ and absolutely convergent near zero for $\text{Re}(\lambda) >0$. From here onward we  use these integral transforms with the understanding that suitable conditions on the parameter $\lambda$ are satisfied to ensure convergence. The value of the momentum wave-function at other values of the momentum is obtained by analytic continuation. 
The inverse transform is given by\footnote{The integral converges for $0<a< 2 \, \text{Re}(j)$. We refer the reader to e.g. \cite{PinkBook} for further details.} 
\be 
f(x) = \frac{1}{2\pi i} \int_{a-i\infty}^{a+i\infty} d\lambda~ \tilde{f}_{\omega}(\lambda)  |x|^{-\lambda}~, \quad\text{with}\quad \omega = \text{sign}(x)~. 
\ee 
For the continuous representation as well, the momenta take values along a shifted imaginary axis. 
%
The momentum eigenstates are then delta-function normalized: 
\begin{align}
\langle \lambda,\omega |\mu, \rho\rangle &=  
2\pi i\,\delta_{\omega \rho}\, \delta(\lambda-\mu)~.
\end{align}
The equality holds for $\text{Re}(\lambda-\mu) =0$. 
%
The completeness relations involve a sum over the two patches:
\begin{align}
    \int_0^{\infty} \frac{dx}{x}~\bigg( |-x\rangle \langle -x|+|x\rangle \langle x|\bigg)&= 1 \\
    \frac{1}{2\pi i} \int_{a-i\infty}^{a+i\infty} d\lambda~ \bigg(|\lambda, -\rangle \langle \lambda, -| + |\lambda, +\rangle \langle \lambda, +|\bigg) &=1 ~. 
\end{align}

\subsection{The Path Integral}

As for the discrete orbits, we now derive matrix elements for the sl$(2,\mathbb{R})$ generators and for the exponentiated operators in the continuous representations using the path integral over the  orbit. 
Once again we work with a discretized path integral in which the total time interval $T$ is divided into $N$ intervals of length $\Delta t$. We introduce phase space variables $(x_a, \pi_a)$ in each interval. The matrix element of an operator ${\cal O}$ between any two momentum states then takes the  form:
\be
\langle \lambda, \omega| {\cal O} | \mu, \rho\rangle = \prod_{a=1}^{N-1} \alpha  \int_{-\infty}^{\infty}\frac{ d\eta_a(t)}{2\pi} \prod_{a=1}^{N}\left( \int_{0}^{\infty} \frac{dx_{a}}{x_{a}}~ e^{iS_{>}} + \int_{0}^{\infty} \frac{d(-x_{a})}{(-x_{a})}~ e^{iS_{<}}\right) ~{\cal O}.
\ee
We make several remarks about the path integral. The variables $\omega$ and $\rho$ take values $\pm$. 
Here we have replaced the integral over the momenta $\pi$ with the integral over the variable $\eta$, which is  defined to be:
\begin{align}
    \eta = \frac{1}{\alpha }Y_2 = \frac{1}{\alpha }(\pi_v - i {j})~.
\end{align}
There is one more position variable than there are  variables. The path in phase space is now chosen such that the initial and final momenta are fixed to be $\mu$ and $\lambda$ respectively. In terms of the $\eta$-variable, the initial and final point in phase space are denoted
\be
\label{etainfin}
\eta_0 = \frac{i}{\alpha}(\mu-{j})~,\quad\text{ and}\quad \eta_{N} = \frac{i}{\alpha}(\lambda-{j})~.
\ee
Note that the $x$-integral runs over both patches and that the action depends on the patch.  Lastly we remark  that the first and last of the  $x_a$-integrals will be over a single patch determined by the labels $(\omega, \rho)$ of the initial and final momentum eigenstates. 
The discretized action is read off from the symplectic structure on the orbit. In one of the patches it is given by
\be
\label{Splus}
S_{>} = -\sum_{a=1}^N \alpha\, \log x_a(  \eta_a  -\eta_{a-1} ) ~. 
\ee
The action $S_{<}$ in the other patch is obtained from the action $S_{>}$ by the replacement $x \rightarrow -x$.

\subsubsection{Kernel for the Hyperbolic Generator}

We begin with the simplest kernel calculation in the path integral formalism, which is the calculation of the  matrix element:
\be
 \langle\lambda, +|Y_2|\mu, +\rangle~. 
\ee
One can think of this as the non-trivial part of the matrix element for the operator $e^{i\Delta t Y_2}$, when the time interval $\Delta t$ is infinitesimal. One then has a single interval and therefore a single $x$-integral to be done. Given that the initial and final patch is chosen to be $(+)$, we see that the path integral localizes onto a single patch. In terms of the phase space variables, we have $Y_2 = \alpha\, \eta$.  Using the action \eqref{Splus}, we obtain the  integral expression for the matrix element: 
\begin{align}
    \langle\lambda, +|Y_2|\mu, +\rangle = \int_0^{\infty} \frac{dx}{x}~ \alpha \eta_0~ e^{i \alpha (\eta_N - \eta_0) \log x }~.
\end{align}
We have chosen the initial value convention while inserting $\eta$ in the path integral. The initial and final $\eta$-values are as in equation \eqref{etainfin} and we therefore obtain
\begin{align}
\label{PIresulty2}
    \langle\lambda, +|Y_2|\mu, +\rangle =i (\mu-{j}) \int_0^{\infty} \frac{dx}{x}~ x^{\mu-\lambda} = 2\pi\, i\, (\mu-{j})\, \delta(\mu-\lambda)~.
\end{align}
The generalization to compute the kernel for a finite time interval is straightforward. 
We wish to compute 
\begin{align}
 K_{++}(\lambda, \mu; e^{i T y_2}) = \frac{1}{2\pi i}  \langle \lambda, +|e^{i T y_2} |\mu, + \rangle~. 
\end{align}
As in our previous path integral analysis for the discrete representation, we include the effect of the operator insertion as an addition to the action. The computational steps are almost identical as in the case of discrete orbits, so we merely present the final form of the discretized action: 
\begin{align}
S_>
&=\alpha\eta_0\log x_1 - \alpha\eta_N\log x_N +  \sum_{a=1}^{N-1} \alpha \eta_a (\log x_{a+1}-\log x_{a} +\Delta t) ~.
\end{align}
We first integrate over the $N-1$ variables $\eta_a$. This leads to the constraint that 
$\log x_{a} = \log x_{a+1} + \Delta t$. We first note that, given the initial and final conditions, this necessarily forces { all} the intermediate $x$'s to lie on the $(+)$ patch. The constraint then eliminates all but one of the $x$-integrals, as we have $\log x_1 = \log x_N + (N-1)\Delta t \equiv \log x_N + T$. Substituting all this into the path integral, we are left with a single $x$-integral given by 
\begin{align}
    K_{++}(\lambda, \mu; e^{i T y_2}) 
    &= \frac{1}{2\pi i} \int_{0}^{\infty}  dx~ x^{\lambda-\mu-1}~ e^{-(\mu-{j})T}= \delta(\lambda-\mu)\,  e^{-(\mu-{j})T}~.
\end{align}
One obtains the identical result for the kernel $K_{--}(\lambda, \mu, e^{i T y_2})$. Moreover, 
given that the  action by $Y_2$ does not change the patch, we find zero for the off-diagonal kernels: $K_{+-}(\lambda, \mu, e^{i T y_2})=0=K_{-+}(\lambda, \mu, e^{i T y_2})$.

\subsubsection{Kernels for the Parabolic Generators}
Next, we compute  the matrix elements for a parabolic generator: 
\be 
\langle\lambda, +|Y_-|\mu, +\rangle~. 
\ee 
The operator $Y_-$ can be written down in terms of the phase space variables as 
\be 
Y_- = x^{-1}({-j} + i\, \alpha\, \eta)~. 
\ee 
For the single interval, the boundary conditions ensure once again that only the action $S_>$ plays a role and we find
\begin{align}
\label{Y-between+}
    \langle\lambda, +|Y_-|\mu, +\rangle &= \int_0^{\infty} \frac{dx}{x} x^{-1} ({-j} + i \alpha \eta_0) x^{\lambda-\mu}\cr
    &=-2\pi i\, \mu\, \delta(\lambda -\mu  -1)~.
\end{align}
For the matrix element of the symmetry generator $Y_-$ between the $(-)$ states, we observe that the main difference is the presence of $x^{-1}$ which is odd under the $x\leftrightarrow (-x)$ exchange. Thus, in the path integral where we use $S_<$ for the action, we pick up an extra sign and we obtain 
\begin{align}
\label{Y-between-}
    \langle\lambda, -|Y_-|\mu, -\rangle &= -\int_0^{\infty} \frac{d(-x)}{(-x)} (-x)^{-1} ({-j} + i \alpha \eta_0) (-x)^{\lambda-\mu}\cr
    &=+2\pi i\, \mu\, \delta(\lambda -\mu  -1)~.
\end{align}
Given the matrix element for the $(+,+)$ overlap, the derivation of the matrix element for the exponentiated $Y_-$ operator proceeds along the same lines as in the discrete case by repeated insertions of the complete set of momentum eigenstates. We  present the result:  
\begin{align}
K_{++}(\lambda, \mu; e^{c Y_-}) &= \frac{1}{2\pi i}\langle \lambda, +| e^{c Y_-}|\mu,+\rangle \cr
&=\frac{1}{2\pi i} \int_0^{\infty}~dx~x^{\lambda-1}(x+c)^{-\mu} 
~. 
\end{align}
For the $(-,-)$ overlap however, due to the flip in sign of the matrix element in \eqref{Y-between-} with respect to the matrix element in \eqref{Y-between+}, the calculation of the kernel (for $c>0$) leads to the result: 
\begin{align}
K_{--}(\lambda, \mu; e^{c Y_-}) &=\frac{1}{2\pi i} \langle \lambda, -| e^{c Y_-}|\mu,-\rangle \cr&= \frac{1}{2\pi i}\sum_{n=0}^{\infty} \frac{c^n}{n!}\int \prod_{j=1}^n \frac{d\mu_j}{2\pi i} \langle\lambda,-| Y_-| \mu_1, -\rangle \prod_{i=1}^{n-1} \langle\mu_i, -| Y_-| \mu_{i+1},-\rangle
\langle\mu_n, -| Y_-| \mu,-\rangle\cr
&=\frac{1}{2\pi i}\sum_{n=0}^{\infty}c^n \frac{\mu(\mu+1)\ldots (\mu+n-1)}{n!} (2\pi i) \delta(\lambda-\mu-n) ~. 
\end{align}
At this point we use the $\delta$-function to replace the $\mu$-factors by $\lambda$-factors. This leads to
\begin{align} 
K_{--}(\lambda, \mu; e^{c Y_-}) &= \frac{1}{2\pi i}\sum_{n=0}^{\infty}c^n \frac{(\lambda-n)(\lambda-n+1)\ldots (\lambda-1)}{n!}(2\pi i) \delta(\lambda-\mu-n)\cr
&=\frac{1}{2\pi i} \int_0^{\infty}\frac{dx}{x} x^{\lambda-\mu}~ (1+\frac{c}{x})^{\lambda-1} \cr
&= \frac{1}{2\pi i}\int_0^{\infty}dx~ x^{-\mu}~(x+c)^{\lambda-1}
~.
\end{align}
The traditional way to write the final $x$-integral is to let the PSL$(2,\mathbb{R})$ act on the initial wavefunction $\langle x, -|\mu \rangle$. A few shifts and sign flips later, one can rewrite the integral in the  form: 
\begin{align}
K_{--}(\lambda, \mu; e^{c Y_-})  
&= \int^{-c}_{-\infty}~dx~(-x)^{\lambda-1}~ (-x-c)^{-\mu} ~.
\end{align}
One can repeat this procedure for the $Y_+$ operator. We omit the details and record the final answers: 
\begin{align}
K_{++}(\lambda, \mu; e^{b Y_+}) &= \langle \lambda, +| e^{b Y_+}|\mu, +\rangle 
= \int_0^{\infty}~\frac{dx}{x}~x^{\lambda}(1+b\, x)^{-2{j}} \left(\frac{x}{1+b\, x}\right)^{-\mu} 
~.\\
K_{--}(\lambda, \mu; e^{b Y_+}) &= \langle \lambda, -| e^{b Y_+}|\mu, -\rangle 
= \int_{0}^{\infty}~ \frac{dx}{x}~ \left(\frac{x}{1+bx}\right)^{\lambda } (1+b\, x )^{2{j}-1} x^{-\mu}
~.
\end{align}
Making the change of variable $x' = - \frac{x}{1+bx}$, one can recast the last kernel in the  form:
\begin{align}
    K_{--}(\lambda, \mu; e^{b Y_+}) &= \int_0^{-\frac{1}{b}}
 \frac{dx}{x} (-x)^{\lambda} (1+bx)^{-2{j}} 
 \left( -\frac{x}{1+bx}\right)^{-\mu}~.
 \end{align}

\subsubsection{The Crossed Kernels}
It is important to note that for the infinitesimal action, we assumed that one never crosses the $x=0$ point and therefore the matrix elements involving a change of patch from $-$ to $+$  were zero. However,  from the final form of the various kernels we have calculated, it is clear that one can read the kernel as the overlap of wavefunctions in which the SL$(2,\mathbb{R})$ group element acts on the initial wavefunction. For instance, for $g=e^{c Y_-}$, we have   
\begin{align}  
K_{++}(\lambda, \mu; g) = \langle \lambda, + | g | \mu, +\rangle  &= \int_{-\infty}^{\infty} \frac{dx}{x} \langle\lambda, +| x\rangle \langle g\cdot x| \mu, +\rangle\cr
&= \int_{-\infty}^{\infty} \frac{dx}{x} x_+^{\lambda}~(g\cdot x)_+^{-\mu} ~,
\end{align}
where the $g$-action in this case is simply given by a translation:
\be 
g\cdot x = x+c~.
\ee 
Therefore, the global picture makes it clear that for a finite transformation, we could have non-vanishing crossed kernels. For instance, 
\begin{align}
   K_{-+}(\lambda, \mu, e^{bY_-})= \langle \lambda, - | e^{bY_-} | \mu, +\rangle  &= \int_{-\infty}^{\infty} \frac{dx}{x}  \langle \lambda, -|x\rangle \langle x | e^{bY_-} | \mu, +\rangle \cr
    &= \int_{-\infty}^{\infty} \frac{dx}{x} x_-^{\lambda} \langle x+b  | \mu, +\rangle\cr
    &= \int_{-\infty}^{\infty} \frac{dx}{x} x_-^{\lambda} (x+b)_{+}^{-\mu}
\end{align}
For a finite $b>0$ we see that it is possible to obtain a non-zero crossed kernel. The kernel $K_{+-}$ can be shown to be zero using the same methods:  
\begin{align}
   K_{+-}(\lambda, \mu, e^{bY_-})= \langle \lambda, + | e^{bY_-} | \mu, -\rangle  &= \int_{-\infty}^{\infty} \frac{dx}{x}  \langle \lambda, +|x\rangle \langle x | e^{bY_-} | \mu, -\rangle \cr
    &= \int_{-\infty}^{\infty} \frac{dx}{x} x_+^{\lambda} \langle x+b  | \mu, -\rangle\cr
    &= \int_{-\infty}^{\infty} \frac{dx}{x} x_+^{\lambda} (x+b)_{-}^{-\mu} = 0~.
\end{align}
The $x_+^\lambda$ term forces the function multiplying it to be evaluated for positive argument alone, and for $b>0$, this is impossible. Thus we obtain zero overlap. 
The kernels for the opposite sign of the parameter are obtained by flipping both the index signs. For instance, keeping $b>0$, we have 
\begin{align}
    K_{++}(\lambda,\mu, e^{-bY_-}) &= K_{--}(\lambda, \mu, e^{bY_-})\\
     K_{+-}(\lambda,\mu, e^{-bY_-}) &= K_{-+}(\lambda, \mu, e^{bY_-})~,
\end{align}
and so on. One can similarly obtain the kernels for the finite $Y_+$ transformation. 

\subsubsection{The s-operator}

From a given patch, the second patch can be reached by acting with the operator $s$. We recall that in PSL$(2,\mathbb{R})$, the group element $s$ squares to one and thus generates a $\mathbb{Z}_2$ subgroup. 
We wish to understand how the group operation $s$ is realized in our path integral representation for the continuous orbit. To that end, we analyze the kernel when such a group element is inserted. 
In the $(x,\eta)$ coordinate system we see that the action of the $s$ transformation flips the sign of the momentum $\eta$. Thus if we consider an infinitesimal path from $\eta_0$ to $\eta_N$, where $\eta_N$ belongs to a different patch, the action corresponds to 
\be 
S = -\alpha\log x (\eta_0 + \eta_N) ~.
\ee 
Integrating over the intermediate position $x$, we thus obtain 
\begin{align}
K_{+-}(\lambda,\mu;s) &= \frac{1}{2\pi i}\int_0^{\infty} \frac{dx}{x} e^{-i\alpha \log x(\eta_0+\eta_N)} \cr
&= \frac{1}{2\pi i} \int_0^{\infty} \frac{dx}{x}  x^{-2 {j} +\lambda+\mu} =\delta(\lambda+\mu -2{j}) ~, 
\nonumber \\
K_{-+}(\lambda, \mu; s) &= 
\frac{1}{2\pi i}\int^{\infty}_{0}\frac{dx}{x}  (-x)^{-2{j}+\mu+\lambda}= \delta(\lambda+\mu -2{j}) \, .
\end{align}
For the same patch kernels we find $K_{++}(\lambda,\mu;s) = 0 = K_{--}(\lambda,\mu;s)$. 
We can summarize the kernel for the group element $s$ as:
\begin{align}
K_{\omega,\rho}(\lambda,\mu;s) &= \frac{1}{2 \pi i}
\int_{-\infty}^{+\infty} x^{\lambda-1}_\omega |x|^{-2 {j}}   \left( -\frac{1}{x} \right)^{-\mu}_\rho dx  \, .
\end{align}
The $s$ operator is realized as the fractional linear transformation $x \rightarrow -1/x$.

\subsubsection{Summary}
If we combine all the results obtained in conjunction with the decomposition \eqref{sl2elementdecomp} of the general PSL$(2,\mathbb{R})$ element, we can write down  
the kernel for the insertion of any   PSL$(2,\mathbb{R})$ element $g={\scriptsize \begin{pmatrix} a & b \\ c & d \end{pmatrix}}$:
\begin{equation}
K_{\omega,\rho}(\lambda,\mu;\chi;g) = \frac{1}{2 \pi i}
\int_{-\infty}^{+\infty} x^{\lambda-1}_\omega |b x + d|^{ -2{j}}  \left( \frac{a x + c}{bx+d} \right)^{-\mu}_\rho dx \, .
\end{equation}
Indeed, this is the  kernel for a continuous PSL$(2,\mathbb{R})$ representation on the Hilbert space of quadratically integrable functions on $\mathbb{R}$. 

Thus the path integration on orbits of both discrete and continuous type gives rise to Hilbert spaces that realize a single irreducible unitary representation of PSL$(2,\mathbb{R})$. We have demonstrated this expectation explicitly in a parameterization adapted to a one-parameter hyperbolic symmetry group. The derivation is laced with surmountable technical hurdles.

\section{The Representations of Covers and Traces}
\label{repsofcovers}
\label{CoveringGroups}
The group SL$(2,\mathbb{R})$ has a standard matrix realization in terms of two-by-two matrices with real elements and unit determinant. The group PSL$(2,\mathbb{R})$ is obtained from this group by dividing by the central group $\mathbb{Z}_2$. We exploited the matrix representation and the fractional linear action of the group on an auxiliary variable in our description of PSL$(2,\mathbb{R}$) kernels above. In this section, we analyze how to extend the kernels  of  PSL$(2,\mathbb{R})$
representations to the covering group SL$(2,\mathbb{R})$ and their universal cover $\widetilde{G}$. The latter has no matrix realization. We will  briefly comment on how one may interpret these extensions  in the path integral formalism.  In order to obtain the generalization, we first formulate a detailed  description of the covering group $\widetilde{G}$. We then describe kernels of representations of SL$(2,\mathbb{R})$ and $\widetilde{G}$. 

The focus of this section however is to use these descriptions to compute the traces of group elements in the covering group $\widetilde{G}$ that project onto hyperbolic group elements  in PSL$(2,\mathbb{R})$. This result will also contain the trace for the restrictions to the subgroups SL$(2,\mathbb{R})$ and PSL$(2,\mathbb{R})$. The trace calculations we perform will be of use in physical applications, e.g. when computing partition functions with hyperbolic insertions representing chemical potentials or other physical quantities of interest.

\subsection{The Universal Covering Group}
\label{DefinitionCoveringGroup}
We have already described the matrix group SL$(2,\mathbb{R})$. We wish to provide a detailed description of the universal covering group $\widetilde{G}$ of SL$(2,\mathbb{R})$. 
To describe the cover, it is useful to pull the group SU$(1,1)$ into our discussion. 
The group SU$(1,1)$ is conjugate to SL$(2,\mathbb{R})$ in GL$(2,\mathbb{C})$. It is isomorphic to SL$(2,\mathbb{R})$. Their covering groups are the same. For some aspects, it can be technically advantageous to work with the group SU$(1,1)$. Indeed, the compact SO$(2)$ subgroup  of SL$(2,\mathbb{R})$  consisting of two-by-two rotation matrices can be thought of as forming a non-trivial loop in SL$(2,\mathbb{R})$ (as explained in subsection \ref{OrbitAndGroup}). The covering group covers this loop as a line covers a circle.  The group SU$(1,1)$ consists of matrices 
\be 
\begin{pmatrix} 
\alpha &\bar{\beta} \\
\beta &\bar{\alpha}
\end{pmatrix} 
\ee 
with $\alpha,\beta$ complex numbers satisfying $|\alpha|^2-|\beta|^2=1$.  The SO$(2)$ subgroup can be realized by diagonal matrices in the SU$(1,1)$ group. This fact gives rise to mild technical simplifications. In more detail, we see this as follows. We parameterize the SU$(1,1)$ group by the coordinates
\be
\gamma=\beta/\alpha \, , \qquad \qquad \omega=\text{arg} \, \alpha \,.
\ee
We have the inverse relations 
\be \alpha=e^{i \omega} (1-|\gamma|^2)^{-\frac12} \, , \qquad \qquad \beta=e^{i \omega} \gamma (1-|\gamma|^2)^{-\frac12} \, .
\ee
The group SU$(1,1)$ is described by the region $|\gamma|<1$ and $- \pi < \omega \le \pi$. The universal cover $\widetilde{G}$ is found by dropping the identification on $\omega$ and taking $\omega \in \mathbb{R}$. The composition law in the cover is defined by the same formulas as in the SU$(1,1)$ composition law for $(\gamma,\omega)$, which in turn follows from  matrix multiplication.
The projection map $\Phi: \widetilde{G} \rightarrow SU(1,1)$ is found by reconsidering the coordinate $\omega$ modulo $2 \pi$. The kernel of the projection is $(\gamma,\omega)=(0, 2 \pi \mathbb{Z})$. 
Part of these observations are summarized in the short exact sequence:
\begin{equation}
0 \rightarrow \mathbb{Z} \rightarrow \widetilde{G} \rightarrow \mathrm{PSL}(2,\mathbb{R}) \rightarrow 1 \, .
\end{equation}
In this sequence, we can replace PSL$(2,\mathbb{R})$ by its double cover SL$(2,\mathbb{R}) \equiv SU(1,1)$.
The center of the universal covering group is the group $\mathbb{Z}$. 

\subsection{The Kernel for SL(2,R)}

The center of the group is realized as a scalar matrix in an irreducible representation. In a unitary representation $r$, a generator
$z$ of the center is therefore represented by a phase factor $r(z)=e^{2 \pi i \epsilon}$ times the unit operator. 
In the sequel, we firstly concentrate on the double cover SL$(2,\mathbb{R})$ of PSL$(2,\mathbb{R})$ and then move to the universal cover $\widetilde{G}$. 
Since $z^2$ is the identity in SL$(2,\mathbb{R})$, we have in that case that $\epsilon=0$ or $\epsilon=1/2$. Whenever we have an insertion of $-e=z$ in the kernel, we add a minus sign if the representation has $\epsilon=1/2$. We already saw in subsection \ref{FirstDiscussionOfMinusE} (or manifestly, in an elliptic basis) that for the discrete representations of SL$(2,\mathbb{R})$, the parity of $2 \epsilon$ equals the parity of $2j$. For the continuous representations, the Casimir and the parity are independent.  
For the group kernel $K$ in the continuous representations, we  find for a $(-e)$ insertion:
\begin{align}
K_{\omega\rho}(\lambda,\mu;{j},\epsilon,-e) &= (-1)^{2 \epsilon}  \frac{1}{2 \pi i}
\int_{-\infty}^{+\infty} x^{\lambda-1}_\omega  x ^{-\mu}_\rho dx 
\end{align}
which says that the kernel is diagonal in momentum space and in $(+,-)$ space and that we pick up an overall sign when the representation represents minus the identity non-trivially. Furthermore, we recall that $s^2=-e$ inside SL$(2,\mathbb{R})$. Thus, we need to make the representation of the operator $s$ consistent with the representation of the operator $-e$.  %
The 
way to implement this consistency condition in accord with the SL$(2,\mathbb{R})$ group multiplication law is:
\begin{align}
K_{\omega,\rho}(\lambda,\mu;{j}, \epsilon, g) &= \frac{1}{2 \pi i}
\int_{-\infty}^{\infty}dx~ x^{\lambda-1}_\omega |b x + d|^{-2 {j}} \text{sgn}^{2 \epsilon} (b x+ d) \left( \frac{a x + c}{bx+d} \right)^{-\mu}_\rho  \, .
\end{align}
We can rewrite the kernel as:
\begin{align}
K_{\omega,\rho}(\lambda,\mu;{j}, \epsilon, g)
&= \frac{1}{2\pi i} \int_{-\infty}^{\infty} \frac{dx}{x}\, x_{\omega}^{\lambda}\int_{-\infty}^{\infty} dx'\, x'^{-\mu}_{\rho}~ |b x + d|^{-2 {j}} \text{sgn}^{2 \epsilon} (b x+ d)\, \delta\left( x'- x\cdot g \right).
\end{align}
This is the kernel of continuous representations realized in the space of quadratically integrable functions on the real line $\mathbb{R}$ parameterized by the variable $x$. The kernel in position space can be read off  by peeling off the Mellin transform: 
\be 
K(x', x; {j},\epsilon, g) =| b x + d|^{-2 {j}}~ \text{sgn}^{2 \epsilon}(b x+ d) ~ \delta\left( x'- x\cdot g \right)~.
\ee 
This kernel is defined such that its action on functions  on the real line is
\begin{align} 
\label{sl2rconttransf}
\left(T_{{j},\epsilon}(g)\cdot f \right)(x) &=\int_{-\infty}^{\infty} dx'~K(x', x;{j}, \epsilon, g)~f(x') \cr
&= | b x + d|^{-2 {j}}~ \text{sgn}^{2 \epsilon}(b x+ d) ~f(x\cdot g)~.
\end{align}

\subsection{The Representations and Kernel of the  Universal Cover}
We move to consider the representations of the universal covering group $\widetilde{G}$ of PSL$(2,\mathbb{R})$. We discuss the main features of the representations, their kernels and their traces. Mathematical treatments of the representations and some  of these results can be found in \cite{Pukanszky,Sally}. We believe the  direct and elementary access we provide to the complete results is useful.

Firstly, the continuous and discrete representations of the universal covering group $\widetilde{G}$ can be defined on the same function spaces as those of $SL(2,\mathbb{R}) \equiv SU(1,1)$.\footnote{The Lie algebra orbits of these groups coincide with those of PSL$(2,\mathbb{R})$.}
Standard function spaces used to represent PSL$(2,\mathbb{R})$  are functions on the upper half plane (or the upper imaginary axis therein) as well as on the real line. See e.g. \cite{PinkBook}. 
The conformal transformation $\phi$ that relates the SL$(2,\mathbb{R})$ group to the SU$(1,1)$ group within GL$(2,\mathbb{C})$ maps the upper half plane to the inside of the unit disk and the real line to the circle. Thus, functions on those spaces  naturally appear as carrying SU$(1,1)$ representations. 
More specifically, the two-by-two matrix that conjugates SL$(2,\mathbb{R})$ to SU$(1,1)$ equals
\be
\phi = \left( \begin{array}{cc}
 1 & -1 \\
 -i & -i  \end{array}
 \right) \, .
 \ee
 A small calculation shows that the SL$(2,\mathbb{R})$ group element $g$ is related to the SU$(1,1)$ $(\alpha_1+i \alpha_2,\beta_1+i \beta_2)$ parameters as
 \be
g = \left( \begin{array}{cc}
 a & b \\
 c & d  \end{array}
 \right) =
 \left( \begin{array}{cc}
 \alpha_1-\beta_1 & -\alpha_2+\beta_2 \\
 \alpha_2+\beta_2 & \alpha_1+\beta_1  \end{array}
 \right)
 \, .
 \ee
 The fractional linear transformation $\phi$ maps the real axis $\mathbb{R}$ into the unit circle $S^1$. The transformation is $z=-i(w-1)/(w+1)$ and if $w$ is on the unit circle, $z$ is real. The map also allows one to write the kernel that implements the SU$(1,1)$ action on functions on the circle: 
 \be 
 \left(T_{s,\epsilon}\cdot f \right)(e^{i\theta}) = \left(\frac{\bar\alpha+e^{i \theta} \bar{\beta}}{\left|\bar\alpha+e^{i \theta} \bar{\beta}\right|} \right)^{2\epsilon} |  \bar{\alpha}+e^{i \theta} \bar{\beta}|^{-1+2is} f(e^{i \theta} \cdot g)~.
 \ee 
 The transformation rule for functions on the circle follows from the fact that the continuous representation of SU$(1,1)$ is unitarily equivalent to the continuous representation of SL$(2,\mathbb{R})$ whose transformation is  in equation \eqref{sl2rconttransf}. We have substituted the continuous representation value ${j}=\frac12- is$ and defined  $w\cdot g= (\alpha w+\beta)/(\bar{\beta} \bar{w} +\bar{\alpha})$ for $w=e^{i\theta}$.

For the continuous representation of the universal covering group $\widetilde{G}$, the action of the group is a subtle extension of this result \cite{Sally}. First of all we recall that $\widetilde{G}=(\gamma,\omega)$ is a group element of the cover as in subsection \ref{DefinitionCoveringGroup} and the projection $\Phi : \widetilde{G} \rightarrow SU(1,1) $ of the covering group element $\tilde{g}$ equals $ \Phi(\gamma,\omega)={\scriptsize \begin{pmatrix} \alpha & \bar{\beta}\\\beta & \bar{\alpha}\end{pmatrix}}=g \in SU(1,1)$. On functions on the  circle the covering group representation is given by \cite{Sally}:
\begin{equation}
(T_{s,\epsilon}(\tilde{g}) f)(e^{i \theta}) = e^{-2 i \omega \epsilon}
\left(\frac{1+e^{i \theta} \bar{\gamma}}{1+e^{-i \theta} \gamma}\right)^\epsilon |e^{i \theta} \bar{\beta} + \bar{\alpha}|^{-1+2is} f(e^{i \theta} \cdot g)~.
\end{equation}
We have rendered the dependence of the transformation law on the phase $\omega$ explicit. 
A crucial point is then that this is actually a representation of the covering group (in which $\omega \in \mathbb{R}$) for all complex values $\epsilon \in \mathbb{C}$ and $s \in \mathbb{C}$ \cite{Sally}. The multiplier phase is carefully isolated in order to  provide a representations of the cover.
We consider unitary continuous representations for which $s \in \mathbb{R}^+$ and $\epsilon \in [0,1[$.\footnote{The case $s=0$ and $\epsilon=1/2$ is an exception we exclude. That representation becomes the direct sum of  two irreducible discrete representations.}

\subsection{Traces}
Before we turn to evaluating the trace of hyperbolic group elements in the representations of the universal cover, we take a step back and review the trace of elliptic group elements.
\subsubsection{The Trace of Elliptic Group Elements}
We recall that we performed the path integral calculation of the trace in section \ref{Elliptic}. The calculation was largely independent of the choice of covering group. The traces of elliptic group elements are still equal to:
\begin{equation}
\text{Tr}^{\pm}_j (e^{i T Y_0}) = \frac{e^{ \pm i j   T}}{1-e^{\pm i   T}} \, ,
\qquad \qquad
\text{Tr}_{\frac{1}{2}+is,\epsilon} (e^{i T Y_0}) = e^{i \epsilon T} \delta_{\mathbb{Z}} (\frac{ T}{2 \pi}) \, .
\end{equation}
The main differences between PSL$(2,\mathbb{R})$ and $\widetilde{G}$ are that the Casimir parameter $j$ is no longer quantized, nor is the parameter $\epsilon$. Both effects are due to the decompactification of the angular direction $\omega$. The trace formula and its path integral derivation remain valid for covering groups.

\subsubsection{The Trace of Hyperbolic Group Elements  in Continuous Representations}
Suppose that we wish to compute the trace of hyperbolic group elements in continuous representations of the covering group of SU$(1,1)$. We set $w=e^{i \theta}$ and find the kernel:
\begin{equation}
K(w,w';g) = 
e^{-2 i \omega \epsilon}
\left(\frac{1+w \bar{\gamma}}{1+\bar{w} \gamma}\right)^\epsilon |w \bar{\beta} + \bar{\alpha}|^{-1+2is} \delta(w \cdot g - w') \, .
\end{equation}
The trace 
\begin{equation}
\text{Tr}\, {T}_{s,\epsilon}(g) = \int_C d w ~ K(w,w;g) 
\end{equation}
of this kernel integrated over the circle $C$
should be thought of as a distribution on the space of functions on the group.
There are fixed points on the circle that satisfy:
\begin{equation}
\frac{\alpha w + \beta}{\bar{\beta} w + \bar{\alpha}}
=w 
\end{equation}
The equation is equivalent to:
\begin{equation}
\label{weqn}
\bar{\beta} w^2 + (\bar{\alpha}-\alpha) w - \beta = 0 \, ,
\end{equation}
after multiplication with $\bar{\beta} w + \bar{\alpha}$ which will appear as a Jacobian factor in the delta-function. We denote the solutions of this equation by $w_k$.
There is a simple relation between the fixed points $w_k$ and the eigenvalues $\lambda_k$ of the SU$(1,1)$ matrix. The eigenvalues  are solutions to the characteristic equation
\be 
\label{su11cp}
\lambda^2-(\alpha+\bar\alpha) + 1 = 0~.
\ee
The two eigenvalues satisfy $\lambda_{1} = \lambda_{2}^{-1}$. By a linear change of variables one can map the eigenvalue equation to the fixed point equation \eqref{weqn}. One can straightforwardly check that the roots of these two equations are related by
\begin{equation}
\label{lambdatoweqn}
\lambda_k = \bar{\alpha} + \bar{\beta} w_k  \, .
\end{equation}
Finally, we can compute the trace:
\begin{align}
\Tr\, {T}_{s,\epsilon}(g) &=e^{-2 i \omega \epsilon} \sum_{k=1}^2
\left(\frac{1+w_k \bar{\gamma}}{1+\bar{w}_k \gamma}\right)^\epsilon |w_k \bar{\beta} + \bar{\alpha}|^{-1+2is}  \left|\frac{(\bar{\beta} w_k+ \bar{\alpha})^2}{
1-
(\bar{\beta} w_k + \bar{\alpha})^2}  \right|
\nonumber \\
&= e^{-2 i \omega \epsilon} \sum_{k=1}^2
\left(\frac{1+w_k \bar{\gamma}}{1+\bar{w}_k \gamma}\right)^\epsilon |w_k \bar{\beta} + \bar{\alpha}|^{2is}  \frac{1}{|\bar{\beta}(w_1-w_2)|}
\, .
\end{align}
We consider the hyperbolic group elements which lie over the diagonal hyperbolic group elements of the group SL$(2,\mathbb{R})$. We set  $\alpha_1=\cosh t$, $\beta_1 = -\sinh t$ and $\alpha_2=0=\beta_2$, for which the fixed points are at $w_1 = 1$ and $w_2=-1$. The eigenvalues of the PSL$(2,\mathbb{R})$ matrix are $e^{\pm t}$. The hyperbolic group elements of the universal cover form  a set which is the direct product of the center $\mathbb{Z}$ and the diagonal hyperbolic elements of PSL$(2,\mathbb{R})$.
This implies that we restrict our group elements to $\omega= n \pi$ where $n \in \mathbb{Z}$. Plugging in the eigenvalues, and taking properly into account the range of parameters, we find the result: 
\begin{align}
\Tr\, {T}_{s,\epsilon}(g) &=
 e^{-2 \pi i n \epsilon}  \frac{e^{2its}+e^{-2its}}{  |e^t-e^{-t}|}
 =
 e^{-2 \pi i n \epsilon}  \frac{\cos 2 ts}{|\sinh t|} \, .
\end{align}

\subsubsection{The Discrete Hyperbolic Trace}
In this subsection, we compute the trace of hyperbolic group elements in discrete representations.\footnote{We improve on a similar  calculation for the group SL$(2,\mathbb{R})$ sketched in \cite{Gurarie}.} Recall that the kernel for the discrete representation $D_j^+$ of SL$(2,\mathbb{R})$ is defined  on functions defined on the positive imaginary semi-axis  -- see the kernel  \eqref{kerneldiscreteony} --:
\begin{equation} 
(T_j(g) f)(i y) =(i\, b\, y +d)^{- 2{{j}}}~f\left(\frac{ i\, a\, y + c}{i\, b\, y + d}\right) ~.
\end{equation}
To perform the calculation, we extend our representation on the positive imaginary axis to the functions on the $z$ upper half plane by defining $z=iy$. As in the case of the continuous representation we  consider the unitarily equivalent representation of the SU$(1,1)$ group and realize the action of the group element $g={\scriptsize \begin{pmatrix} \alpha & \bar\beta \\ \beta & \bar\alpha\end{pmatrix}}$ on those functions: 
\begin{equation}
(T_j(g) f)(z) =  (\bar\alpha+ \bar{\beta} z)^{-2j} f (z \cdot g) \, .
\end{equation}
This presents a natural starting point to extend the representation to one of the universal cover. We consider the  parameter $j$ to be a real number bigger than $1/2$ and rewrite the $(\alpha, \beta)$ variables in terms of the coordinates $(\gamma, \omega)$ that parameterize the  cover. As explained in Section \ref{DefinitionCoveringGroup} this is done by extending $\omega$ to be an arbitrary real number. The covering group is then represented on the functions in the upper half plane as \cite{Sally}: 
\begin{equation}
(T_j(\tilde{g}) f)(z) = e^{2 i \omega j} (1-|\gamma|^2)^{j} (1+ \bar{\gamma} z)^{-2j} f (z \cdot g) \, .
\end{equation}
The calculation of the trace of the hyperbolic element is subtle. We introduce the integral transform to functions of momentum $p$:\footnote{The integral transform is associated to a choice of basis in which one diagonalizes a parabolic generator.}
\begin{align}
F(z) &= \int_0^\infty\, dp\, e^{i pz} \, f(p)~,
\end{align}
and its inverse 
\begin{align}
f(p) &= \frac{1}{2 \pi} \int_{C_z}dz \, e^{-ipz}\, F(z)~ . \label{ParabolicTransform}
\end{align}
The choice of contour $C_z$ must be such that the integral converges and it must be close to the upper half plane where the function $F(z)$ is well-defined. 
The kernel for the representation in the momentum basis can be obtained by doing the integral transform and we formally obtain:
\begin{equation}
K(p,p'; j, g) = \frac{1}{2 \pi} 
    e^{2 i \omega j}~ (1-|\gamma|^2)^j\int_{C_z} dz \,  (1+ \bar{\gamma} z)^{-2j}  e^{i (p'z_g-pz)}  \, ,
\end{equation}
where we have introduced the notation $z_g = z\cdot g $.
To further guide our calculation, it is instructive to have a concrete hyperbolic group element in mind. We fix $\alpha=\cosh t$ and $\beta=-\sinh t$ with $t>0$ for that purpose. We shall return to the general case in due course. 
In the case at hand, $\bar{\gamma}=-\tanh t$ is negative and the diagonal kernel has a singularity for positive real $z=\cosh t$ beyond those created at the fixed points at $z=\pm 1$ where $z_g=z$. To avoid the singularities and to remain close to the upper half plane, we choose the  $z$ contour to be the real axis slightly rotated in the clockwise direction. In this manner, when we close the $C_z$ contour  in the lower half plane (as suggested by the exponent in \eqref{ParabolicTransform}) we avoid a potential contribution from the spurious singularity at $z=\coth t$. A consequence is that the fixed point at $z=1$ will not contribute. We do expect a contribution from the fixed point $z=-1$ which lies below the contour. Thus, in the definition of the kernel, we propose that the rotated contour $C_z$ is given by
\be 
\text{Im}(z) + \epsilon \, \text{Re}(z) = 0~,\ee 
where $\epsilon>0$. 
The  trace will be computed by setting $p'=p$ and integrating over the momentum:
\begin{align}
\Tr\, T_j(\tilde g) &= \int_{C_p} dp~ K(p,p;j,g) \, .
\end{align}
We will perform the $p$-integral by rotating the $p$ integration domain to counterbalance the rotation of the $z$ contour. Despite the fact that we compensate $z$ and $p$ contour rotations, the  operation is non-trivial because it fixes the relative positions of the contour $C_z$ and the fixed points. Next, we
 perform a change of variables (similar to equation \eqref{lambdatoweqn}): 
\begin{align}
\lambda &= \bar{\alpha} + \bar{\beta} z
\, .
\end{align}
The diagonal kernel then takes the form 
\begin{align}
 K(p,p;j,g) 
    &= \frac{\bar\alpha^{2j}}{2 \pi \bar{\beta}} 
    e^{\frac{i p}{\bar\beta}(\alpha+\bar\alpha) }
    e^{2 i \omega j}~ (1-|\gamma|^2)^j~ \int_{C_\lambda}  d\lambda~ \lambda^{-2j}  e^{-\frac{i p}{\bar\beta}(\frac{1}{\lambda}+\lambda) }
    \, .
\end{align}
We further map $\lambda \rightarrow \lambda^{-1}$, which maps the contour to a circle $C'$ in the $\lambda$ plane.
The kernel turns into
\begin{align}
K(p,p;j,g) &=\frac{\bar\alpha^{2j}}{2 \pi \bar{\beta}} 
    e^{\frac{ip}{\bar\beta}(\alpha+\bar\alpha) }
    e^{2 i \omega j}~ (1-|\gamma|^2)^j~ \oint_{C'}  d\lambda~ \lambda^{2j-2}  e^{-\frac{ip}{\bar\beta}(\frac{1}{\lambda}+\lambda) }
    \, .
\end{align}
Finally, we compute the trace: 
\begin{align}
\Tr\, T_j(\tilde g)&=\frac{\bar\alpha^{2j}}{2 \pi \bar{\beta}} 
  e^{2 i \omega j}~ (1-|\gamma|^2)^j~ \oint_{C'}  d\lambda~ \lambda^{2j-2}\int_{C_p} dp~ e^{\frac{i p}{\bar\beta}(\alpha+\bar \alpha -\lambda- \frac{1}{\lambda})}~.  \nonumber \\
&=
\frac{\bar\alpha^{2j}}{2 \pi i } 
  e^{2 i \omega j}~ (1-|\gamma|^2)^j~ \oint_{C'}  d\lambda~  \frac{\lambda^{2j-1}}{\lambda^2-\lambda(\alpha+\bar\alpha)+1}~. 
    \end{align}
The $p$-integral was performed by analytic continuation in the exponent.
We recognize the denominator as the characteristic polynomial of the SU$(1,1)$ matrix (see \eqref{su11cp}). In these variables, we pick up the smaller of the roots of the characteristic polynomial which equals $\lambda=e^{-t}$ (since we chose $t>0$ initially). Note that this indeed corresponds to the $z=-1$ fixed point after the double change of variables, as foreshadowed above.  Combining this observation with the fact that the angle $\omega$ can differ from the angle of $\bar{\alpha}$ by an integer multiple of $2\pi$ and 
denoting by $m$ the copy of SL$(2,\mathbb{R})$  that we are in, we find the result:
\begin{equation}
\text{Tr}\, T_j(\tilde g)  = e^{4\pi i m j}
e^{2 \pi i n j}
\frac{e^{-(2j-1)|t|}}{|\sinh t|} \, .
\end{equation}
In our final result we have added two extensions. The first is the result for the case $t<0$. In both cases, it is the smaller eigenvalue that winds up in the trace. This is the origin of the absolute value sign in the result. Secondly, we have added the dependence of the trace on the sign of the eigenvalue which we have indicated with a 
number $n$, which equals one for a negative eigenvalue and zero for a positive eigenvalue.  The further generalization to the discrete minus representation $D_j^-$ is obtained similarly and reads:
\begin{equation}
\text{Tr}\, T_j^\pm(\tilde g)  = e^{\pm 2\pi i (2m+n) j}
\frac{|\lambda_s|^{2j-1}}{|\lambda_s-\lambda_s^{-1}|} \, ,
\end{equation}
where $|\lambda_s|$ is the smaller absolute value of the two eigenvalues. Since our derivation is based on a conjectural identification of the contour of integration in the trace of the kernel, it is useful to check our final formula against the literature. 
On the one hand, the sum of the characters of two discrete representations $D^+_j$ and $D^-_j$ was computed in \cite{Pukanszky} and matches the sum of our results. 
On the other hand, for the ordinary group SL$(2,\mathbb{R})$, our result agrees with the formula proposed in \cite{PinkBook,Gurarie} for individual discrete representations.  This is ample confirmation for the contour prescription. Thus, our expression for the trace of hyperbolic group elements in discrete representations reliably extends these results to the individual representations of the covering group. 

\subsection{The Path Integral Perspective}
The adjoint orbit is an orbit for all groups, from PSL$(2,\mathbb{R})$ to its universal cover $\widetilde{G}$. 
Central elements in the group $G$ that we wish to represent leave the orbit invariant. Thus, a path integral quantization will proceed from the same basic ingredients as for PSL$(2,\mathbb{R})$. We do need to define the path integral with central element insertions separately. A priori, we know that the center of the group $G$ is represented by phases since the group action is transitive on the orbit (or the representation is irreducible). Moreover,  we can associate central elements with  non-trivial loops in the group PSL$(2,\mathbb{R})$ and gather from this that the action of the central element is reflected in a rotation of the orbit, as explained in detail in subsection \ref{OrbitAndGroup}. The transformation of the world line action by this rotation was discussed in subsection \ref{FirstDiscussionOfMinusE} and reveals the representation dependent phase.

Effectively then, this definition of the path integral on covering groups $G$ of PSL$(2,\mathbb{R})$ will be equivalent to the properties of the kernels introduced above. We neither loose nor gain much by treating the representations of covering groups in this manner from a path integral perspective. The difference lies in overall phases.

\subsubsection{Summary} 
In this section, we presented kernels of representations of covering groups in function spaces. We used them to compute the trace of elliptic and hyperbolic group elements in the discrete and continuous representations. The calculation of the traces of group elements in representations of non-compact groups is more subtle than in compact groups, and it is equally useful in physical applications.

\section{Conclusions and Perspectives}
\label{Conclusions}
We studied the path integral on coadjoint orbits of $sl(2,\mathbb{R})$.  We stressed that in the case of the group PSL$(2,\mathbb{R})$, there are multiple natural slicings of the orbit corresponding to one-parameter subgroups of elliptic, hyperbolic or parabolic nature. The elliptic slicing of the orbits gives rise to an extension of the orbit theory on compact Lie groups.  Our focus was on carrying out the path integral quantization of the orbits in coordinates adapted to a hyperbolic one-parameter subgroup action on the orbits. This quantization exhibits many new features, such as a continuous spectrum of eigenvalues, and the more subtle treatment of the global geometry of the orbit. The path integral allows for the explicit calculation of the action of the group in an irreducible unitary representation. We used the quantization to compute the traces of elliptic and hyperbolic group elements in continuous and discrete representations of (covering groups of)  $\mathrm{PSL}(2,\mathbb{R})$. 

We view our results as having an interest in their own right, as an elementary contribution to mathematical physics. We also believe these are stepping stones towards solving other problems.

 One  direction for future research is to extend our path integral treatment to hyperbolic slices of orbits of other non-compact (e.g. simple) Lie algebras (including $so(1,n)$ and $so(2,n)$).  Another generalization lies in studying more complicated physical models with $sl(2,\mathbb{R})$ symmetry. In particular, one can view the quantization of orbits as the most elementary building block in the quantization of particles on group manifolds. Those models can be generalized to the quantization of strings on group manifolds. Those models in turn can be extended to quantizing strings on orbifolds of group manifolds. This sequence of generalizations is relevant to the following  problem. There is a three-dimensional (BTZ) black hole space-time which is an orbifold of a group manifold \cite{Banados:1992gq}. Indeed, the black hole is  a $\mathbb{Z}$ orbifold of the universal covering group of SL$(2,\mathbb{R})$. The latter coincides with an $AdS_3$ space-time manifold.  Moreover, the orbifold  group $\mathbb{Z}$  is embedded into a left-right hyperbolic one-parameter subgroup of the isometry group of $AdS_3$ \cite{Banados:1992gq}. The natural  coordinate system  and basis in which to study the BTZ orbifold are the hyperbolic ones. The calculation of the traces in this paper are a useful preparation for computing partition functions for a particle or string on a BTZ background with insertions of the hyperbolic energy and angular momentum operators. Indeed, we hope to put the techniques we developed here to use in that context among others.

Independently, we believe our contribution adds welcome physical  insight to  basic constructions in special function theory related to elementary non-compact Lie group theory and to the wealth of physical models closely related to compact and non-compact Lie groups.


\bibliographystyle{JHEP}

\begin{thebibliography}{99}
\bibitem{Kostant}B.~Konstant, ``Quantization and Unitary Representations,'' Lectures in
Modern Analysis and Applications III, Lecture Notes in Mathematics, Vol.
170, Springer, Berlin, Heidelberg, 1970.
\bibitem{Souriau}J.-M.~Souriau, ``Structure des systèmes dynamiques,'' Maîtrises de mathématiques, Dunod, 1970.
\bibitem{Kirillov}
A.~Kirillov,``Lectures on the orbit method,'' American Mathematical Soc., Vol. 64, 2004.
\bibitem{Nielsen:1987sa}
H.~Nielsen and D.~Rohrlich,
``A Path Integral to Quantize Spin,''
Nucl. Phys. B \textbf{299} (1988), 471-483
doi:10.1016/0550-3213(88)90545-7

\bibitem{Johnson:1988qm}
K.~Johnson,
``Functional Integrals for Spin,''
Annals Phys. \textbf{192} (1989), 104
doi:10.1016/0003-4916(89)90120-6
\bibitem{Alekseev:1988vx}
A.~Alekseev, L.~Faddeev and S.~Shatashvili,
``Quantization of symplectic orbits of compact Lie groups by means of the functional integral,''
J. Geom. Phys. \textbf{5} (1988), 391-406
doi:10.1016/0393-0440(88)90031-9
\bibitem{PinkBook}
N.~~Vilenkin, A.~~Klimyk, ``Representation of Lie Groups and Special Functions: Volume 1: Simplest Lie Groups, Special Functions and Integral Transforms,'' Mathematics and its Applications, 72, 1991.
\bibitem{Vergne}
M.~Vergne,  ``Representations of Lie groups and the orbit method,'' Emmy Noether in Bryn Mawr. Springer, New York, NY, 1983. 59-101.
\bibitem{Witten:1987ty}
E.~Witten,
``Coadjoint Orbits of the Virasoro Group,''
Commun. Math. Phys. \textbf{114} (1988), 1
doi:10.1007/BF01218287
\bibitem{Troost:2003ge}
J.~Troost and A.~Tsuchiya,
``Three-dimensional black hole entropy,''
JHEP \textbf{06} (2003), 029
doi:10.1088/1126-6708/2003/06/029
[arXiv:hep-th/0304211 [hep-th]].
\bibitem{Troost:2012ck}
J.~Troost,
``Models for modules: The story of O,''
J. Phys. A \textbf{45} (2012), 415202
doi:10.1088/1751-8113/45/41/415202
[arXiv:1202.1935 [hep-th]].
\bibitem{Pukanszky}
L. Pukanszky, ``The Plancherel formula for the universal covering group of SL(R, 2)'', Math.
Annalen 156, 96–143 (1964).

\bibitem{Sally}
P.~Sally, ``Analytic Continuation of the Irreducible Unitary Representations of the Universal Covering Group of SL(2, R),''
Issue 69 of Memoirs of the American Mathematical Society, American Mathematical Soc., 1967.
\bibitem{Gurarie}
D.~Gurarie, ``Symmetries and Laplacians: introduction to harmonic analysis, group representations and applications,'' Courier Corporation, 2007.
\bibitem{Banados:1992gq}
M.~Banados, M.~Henneaux, C.~Teitelboim and J.~Zanelli,
``Geometry of the (2+1) black hole,''
Phys. Rev. D \textbf{48} (1993), 1506-1525
[erratum: Phys. Rev. D \textbf{88} (2013), 069902]
doi:10.1103/PhysRevD.48.1506
[arXiv:gr-qc/9302012 [gr-qc]].

\end{thebibliography}

\end{document}